\documentclass[12pt,a4paper,titlepage,onecolumn,final,oneside]{book}
\usepackage[a4paper,left=3cm,right=2cm,top=3cm,bottom=2cm]{geometry}
\usepackage[utf8]{inputenc}
\usepackage{amsmath}
\usepackage[brazil]{babel}
\usepackage{setspace}
\usepackage[round]{natbib} 
\usepackage{lastpage}	
\usepackage{subfigure} 	
\usepackage[font=small,font=singlespacing,format=plain]{caption}	
\DeclareCaptionLabelSeparator{colon}{ - }
\usepackage{rotating}
\usepackage{indentfirst}
\usepackage{fancyhdr}	
\usepackage{titlesec}	
\titlespacing{\chapter}{0pt}{-5pt}{0pt}
\titlespacing{\section}{0pt}{0pt}{0pt}
\titlespacing{\subsection}{0pt}{0pt}{0pt}
\titleformat{\chapter}[block]
{\normalfont\bfseries}{\thechapter}{}{}
\titleformat{\section}
{\normalfont\bfseries}{\thesection}{1em}{}
\titleformat{\subsection}
{\normalfont}{\thesubsection}{1em}{}
\titleformat{\subsubsection}
{\normalfont}{\thesubsubsection}{1em}{}
\addto\captionsbrazil{%
}
\usepackage[titles,subfigure]{tocloft}	
\usepackage[final,breaklinks]{hyperref}		                   
\hypersetup{colorlinks=true, linkcolor=blue, citecolor=blue, filecolor=blue, urlcolor=blue, pdfauthor={José Henrique Hildebrand Grisi Filho}, pdftitle={Caracteriza\c{c}\~{a}o de circuitos pecu\'{a}rios com base em redes de movimenta\c{c}\~{a}o de animais}, pdfproducer={Latex}}	
\usepackage{graphicx}
\usepackage{type1cm}
\usepackage{eso-pic}
\usepackage{color}

\makeatletter
\DeclareRobustCommand{\em}{%
  \@nomath\em \if b\expandafter\@car\f@series\@nil
  \normalfont \else \bfseries \fi}
\makeatother

\author{José Henrique Hildebrand Grisi Filho}
\title{Caracteriza\c{c}\~{a}o de circuitos pecu\'{a}rios com base em redes de movimenta\c{c}\~{a}o de animais}
\date{2012}

\begin{document}
\fancypagestyle{plain}{%
    \fancyhf{} 
    \fancyhead[R]{\thepage} 
    \renewcommand{\headrulewidth}{0pt} 
}
\thispagestyle{empty}
\begin{center}
{\Large José Henrique Hildebrand Grisi Filho}
\par
\vspace{200pt}
{\LARGE Caracteriza\c{c}\~{a}o de circuitos pecu\'{a}rios com base em redes de movimenta\c{c}\~{a}o de animais}
\par
\vspace{377pt}
\textbf{{\large São Paulo}\\
{\large 2012}}
\end{center}
\clearpage{\pagestyle{empty}\cleardoublepage}
\thispagestyle{empty}

\begin{center}
{\Large José Henrique Hildebrand Grisi Filho}
\par
\vspace{100pt}
{\LARGE Caracteriza\c{c}\~{a}o de circuitos pecu\'{a}rios com base em redes de movimenta\c{c}\~{a}o de animais}
\end{center}
\par
\vspace{50pt}
\hspace*{16em}\parbox{10cm}{{Tese apresentada ao Programa de Pós-Graduação em Epidemiologia Experimental Aplicada às Zoonoses
da Faculdade de Medicina Veterinária e Zootecnia da Universidade de São Paulo para obtenção do título de Doutor em Ciências}}
\par
\vspace{50pt}
\hspace*{16em}\parbox{10cm}{{\textbf{Departamento:}\\Medicina Veterinária Preventiva e Saúde Animal}}
\par
\vspace{50pt}
\hspace*{16em}\parbox{10cm}{{\textbf{Área de concentração:}\\Epidemiologia Experimental Aplicada às Zoonoses}}
\par
\vspace{50pt}
\hspace*{16em}\parbox{10cm}{{\textbf{Orientador:} Prof. Dr. Marcos Amaku}}
\par
\vspace{40pt}
\hspace*{16em}\parbox{10cm}{{De acordo: \line(1,0){100}}}
\hspace*{23em}\parbox{10cm}{{Orientador(a)}}
\par
\vspace{50pt}
\begin{center}
\textbf{{\large São Paulo}\\
{\large 2012}}\\
\vspace{10pt}
\textbf{Obs.: A versão original se encontra disponível na Biblioteca da FMVZ/USP}
\end{center}
\makeatletter
\AddToShipoutPicture*{%
            \setlength{\@tempdimb}{.5\paperwidth}%
            \setlength{\@tempdimc}{.5\paperheight}%
            \setlength{\unitlength}{1pt}%
            \put(\strip@pt\@tempdimb,\strip@pt\@tempdimc){%
        \makebox(0,0){\rotatebox{30}{\textcolor[gray]{0.75}%
        {\fontsize{2cm}{2cm}\selectfont{VERSÃO CORRIGIDA}}}}%
            }%
}
\makeatother
\newpage
\clearpage{\pagestyle{empty}\cleardoublepage}
\thispagestyle{empty}


\newpage
\begin{center}
{\textbf{FOLHA DE AVALIAÇÃO\\}}
\end{center}
{\raggedright
Nome: Grisi Filho, José Henrique Hildebrand\par
Título: Caracteriza\c{c}\~{a}o de circuitos pecu\'{a}rios com base em redes de movimenta\c{c}\~{a}o de animais\\}
\begin{center}
\par
\vspace{30pt}
\hspace*{16em}\parbox{10cm}{{Tese apresentada ao Programa de Pós-Graduação em Epidemiologia Experimental Aplicada às Zoonoses
da Faculdade de Medicina Veterinária e Zootecnia da Universidade de São Paulo para obtenção do título de Doutor em Ciências}}\\
\vspace{20pt}
{\raggedright
\textbf{Data:\line(1,0){30}/\line(1,0){30}/\line(1,0){30}}\\}
\vspace{20pt}
{\textbf{BANCA EXAMINADORA}}
\vspace{2em}
\begin{center}
\begin{minipage}{15cm}
\large 
Prof. Dr.  :\line(1,0){363}\\
Instituição:\line(1,0){180}\hspace{1pt} Julgamento:\line(1,0){100}\par \vspace{20pt}
Prof. Dr.  :\line(1,0){363}\\
Instituição:\line(1,0){180}\hspace{1pt} Julgamento:\line(1,0){100}\par \vspace{20pt}
Prof. Dr.  :\line(1,0){363}\\
Instituição:\line(1,0){180}\hspace{1pt} Julgamento:\line(1,0){100}\par \vspace{20pt}
Prof. Dr.  :\line(1,0){363}\\
Instituição:\line(1,0){180}\hspace{1pt} Julgamento:\line(1,0){100}\par \vspace{20pt}
Prof. Dr.  :\line(1,0){363}\\
Instituição:\line(1,0){180}\hspace{1pt} Julgamento:\line(1,0){100}\par \vspace{20pt}
\end{minipage}
\end{center}
\end{center}
\newpage

\pagestyle{empty} \section*{Agradecimentos}
\par\vspace*{50pt} Ao Prof. Dr. Marcos Amaku, pela orientação, sabedoria, paciência e confiança;
\par\vspace*{20pt} Ao Prof. Dr. Fernando Ferreira, por toda a ajuda, sugestões, apoio e amizade;
\par\vspace*{20pt} Aos colegas Raul e Rísia, pelo trabalho em equipe sem o qual esta tese não seria a mesma;
\par\vspace*{20pt} À todos os colegas do LEB que fazem deste um excelente local de trabalho;
\par\vspace*{20pt} À minha família, por tudo;
\par\vspace*{50pt} Projeto financiado pela FAPESP e CNPq.
\par\vspace*{20pt} Ao Instituto de Defesa Agropecuária do Estado do Mato Grosso, pela disponibilização de dados essenciais ao trabalho.
\par\vspace*{50pt} À todos que contribuíram, de forma direta ou indireta, para a conclusão desta tese.

\newpage
\setlength{\parindent}{50pt}
\setlength{\parskip}{1.5ex}
\chapter*{RESUMO}
\thispagestyle{empty}
\begin{singlespace} \noindent
GRISI-FILHO, J. H. H. \textbf{Caracteriza\c{c}\~{a}o de circuitos pecu\'{a}rios com base em redes de movimenta\c{c}\~{a}o de animais.} [Characterization of production zones based on animal movement networks.] 2012. \pageref{LastPage} f. Tese (Doutorado em Ciências) - Faculdade de Medicina Veterinária e Zootecnia, Universidade de São Paulo, São Paulo, 2012.
\end{singlespace}
\begin{doublespace} \noindent
Uma rede é um conjunto de nós conectados entre si por um conjunto de arestas. Redes podem representar qualquer conjunto de objetos que possuam relações entre si.
Comunidades s\~{a}o conjuntos de n\'{o}s relacionados de uma maneira significativa, provavelmente compartilhando propriedades e/ou atuando de forma similar dentro de uma rede. 
Quando a análise de redes é aplicada ao estudo de padrões de movimentação animal, as unidades epidemiológicas de interesse (propriedades, estabelecimentos, municípios, estados, países, etc) são representadas como nós, enquanto a movimentação animal entre elas é representada por arestas de uma rede.
Descobrir a estrutura de uma rede, e portanto as prefer\^{e}ncias e rotas comerciais, pode ser \'{u}til para um pesquisador ou gestor de sa\'{u}de animal. 
Foi implementado um algoritmo de detec\c{c}\~{a}o de comunidades para encontrar grupos de propriedades que \'{e} consistente com a defini\c{c}\~{a}o de circuito pecu\'{a}rio, assumindo que uma comunidade \'{e} um grupo de nós (fazendas, abatedouros) no qual um animal vai mais provavelmente permanecer durante sua vida. 
Este algoritmo foi aplicado na rede interna de movimenta\c{c}\~{a}o animal de 2007 do Estado do Mato Grosso. Esse banco de dados cont\'{e}m informa\c{c}\~{a}o sobre 87.899 propriedades e 521.431 movimenta\c{c}\~{o}es durante o ano, totalizando 15.844.779 de animais movimentados. 
O algoritmo de detec\c{c}\~{a}o de comunidades encontrou uma parti\c{c}\~{a}o da rede que mostra um claro padr\~{a}o geogr\'{a}fico e comercial, duas importantes caracter\'{i}sticas para aplica\c{c}\~{o}es em medicina veterin\'{a}ria preventiva, al\'{e}m de possuir uma interpreta\c{c}\~{a}o clara e significativa em redes de com\'{e}rcio onde liga\c{c}\~{o}es se estabelecem a partir da escolha dos nós envolvidos.\\
Palavras-chave: Epidemiologia. Análise de redes. Análise de comunidades. Circuito pecuário. Movimentação animal.
\end{doublespace}
\chapter*{ABSTRACT}
\thispagestyle{empty}
\begin{singlespace}\noindent
GRISI-FILHO, J. H. H. \textbf{Characterization of production zones based on animal movement networks.} [Caracteriza\c{c}\~{a}o de circuitos pecu\'{a}rios com base em redes de movimenta\c{c}\~{a}o de animais.] 2012. \pageref{LastPage} f. Tese (Doutorado em Ciências) - Faculdade de Medicina Veterinária e Zootecnia, Universidade de São Paulo, São Paulo, 2012.
\end{singlespace}
\begin{doublespace}\noindent
A network is a set of nodes that are linked together by a set of edges. Networks can represent any set of objects that have relations among themselves.
Communities are sets of nodes that are related in an important way, probably sharing common properties and/or playing similar roles within a network. 
When network analysis is applied to study the livestock movement patterns, the epidemiological units of interest (farm premises, counties, states, countries, etc.) are represented as nodes, and animal movements between the nodes are represented as the edges of a network.
Unraveling a network structure, and hence the trade preferences and pathways, could be very useful to a researcher or a decision-maker.
We implemented a community detection algorithm to find livestock communities that is consistent with the definition of a livestock production zone, assuming that a community is a group of farm premises in which an animal is more likely to stay during its life time than expected by chance.
We applied this algorithm to the network of within animal movements made inside the State of Mato Grosso, for the year of 2007. This database holds information about 87,899 premises and 521,431 movements throughout the year, totalizing 15,844,779 animals moved.
The community detection algorithm achieved a network partition that shows a clear geographical and commercial pattern, two crucial features to preventive veterinary medicine applications, and also has a meaningful interpretation in trade networks where links emerge from the choice of trader nodes.\\
Keywords: Epidemiology. Network analysis. Community analysis. Livestock production zone. Animal movement.

\tableofcontents
\thispagestyle{empty}
\pagestyle{myheadings}

\chapter*{1 INTRODUÇÃO}
\addcontentsline{toc}{chapter}{1 INTRODUÇÃO}\stepcounter{chapter}
\markboth{}{}
\doublespacing
\section{Redes}
\begin{figure}[h]
\centering
\caption[Exemplo de rede não-direcionada]{Ilustração de uma rede, com indicação de nós e arestas}
\includegraphics[scale=0.4]{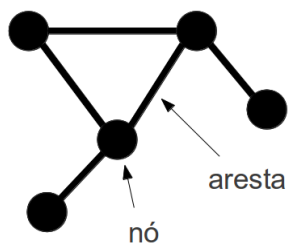}
\label{fig:exemplo1}
\end{figure}

\begin{figure}[h]
\centering
\caption[Rede de associações entre golfinhos]{Rede de associações frequentes entre 62 golfinhos de Doubtful Sound, New Zealand \citep{Lusseau2003a}}
\includegraphics[scale=0.4]{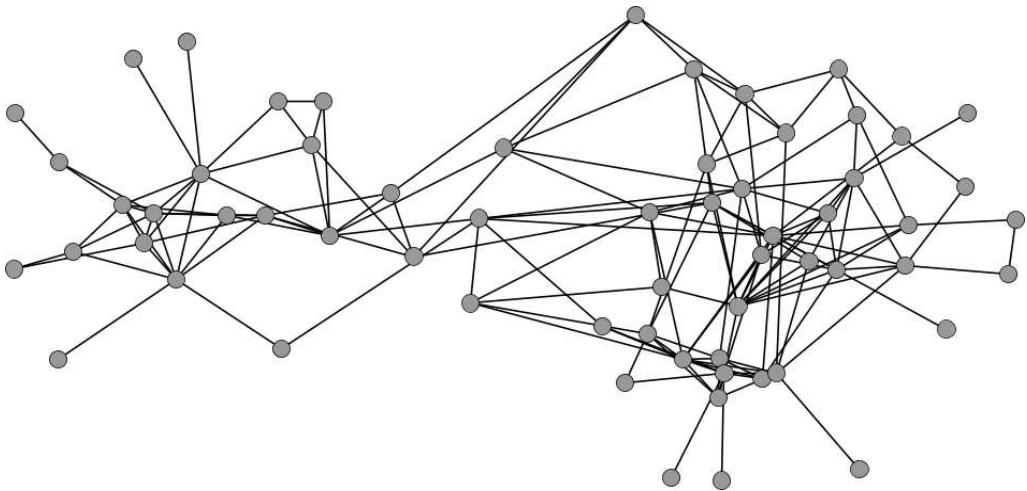}
\label{fig:dolphins2003}
\end{figure}

Uma rede é um conjunto de nós (ou vértices) conectados entre si por um conjunto de arestas (ou ligações), como demonstrado na figura~\ref{fig:exemplo1}. Redes podem representar qualquer conjunto de objetos que possuam relações entre si. Podem, por exemplo, representar as relações sociais entre indivíduos (figura~\ref{fig:dolphins2003}). Em uma rede ponderada cada aresta possui um valor, que representa a intensidade da ligação entre dois nós. Redes podem ser direcionadas e não-direcionadas: uma rede direcionada apresenta a direção da relação entre dois nós, sendo portanto uma ligação de saída para um e uma ligação de entrada para o outro; uma rede não-direcionada assume que a relação entre dois nós é sempre recíproca. Em uma rede direcionada e ponderada, o grau de entrada de um nó ($k_i^{in}$) é o número de nós conectados a este por ligações de entrada, enquanto o grau de saída ($k_i^{out}$) é o número de nós conectados a este por ligações de saída. Analogamente, o grau ponderado de entrada ($kw_i^{in}$) é a soma dos valores das arestas que um nó recebe, e o grau ponderado de saída ($kw_i^{out}$) é a soma dos valores das arestas que um nó envia. O grau ponderado total ($kw_i$) é igual à soma dos graus ponderados de entrada e de saída, enquanto o grau total ($k_i$) é o número de nós ligados ao nó de interesse, independente da direção da ligação.
\par Quando a análise de redes é aplicada ao estudo da movimentação animal dentro de um sistema produtivo, as unidades epidemiológicas de interesse (propriedades, estabelecimentos, municípios, estados, países, etc) são representadas como nós, enquanto a movimentação animal entre elas é representada por arestas de uma rede, onde o valor de cada aresta traz o número de animais comercializados entre dois nós. Assim, o grau de entrada de um nó equivale à quantidade de unidades que enviaram animais para este e o grau de saída equivale à quantidade de estabelecimentos para os quais um nó enviou animais. O grau ponderado de entrada equivale à quantidade de animais comprados e o grau ponderado de saída equivale à quantidade de animais vendidos por uma propriedade.
A análise de redes têm sido utilizada para estudar uma grande variedade de problemas dentro da medicina veterinária, incluindo análises de epidemias \citep{Gibbens2001}, análises etológicas \citep{Lusseau2004}, padrões de distribuição de doenças, modelos preditivos \citep{Harvey2007b}, análises de risco, eficácia de sistemas de vigilância entre outros. \citet{Martinez-Lopez2009} fazem uma revisão sobre o assunto.
\section{Análise de Comunidades em Medicina Veterinária}
Comunidades são conjuntos de nós relacionados entre si de uma maneira significativa, provavelmente compartilhando propriedades e/ou atuando de forma similar dentro de uma rede. Estudos recentes utilizaram a análise de comunidades em redes de trânsito animal com o intuito de revelar a estrutura subjacente da rede \citep{Green2009a,Green2011,Lentz2011}.
\par Descobrir essa estrutura e portanto as preferências de comércio entre propriedades pode se mostrar útil para um pesquisador ou tomador de decisão.
Em primeiro lugar, permite a identificação dos fluxos de comércio entre zonas de produção, possibilitando um melhor entendimento do sistema de produção animal. A existência e funcionalidade de grupos dentro de uma rede pode influenciar o espalhamento de doenças, alterando a velocidade de espalhamento, o tamanho final e a duração de uma epidemia \citep{Jones2010}. Análises de comunidade poderiam ser usadas para determinar os grupos de uma amostra estratificada que visa levar em consideração o trânsito animal, ou ajudar no planejamento de um sistema de vigilância voltado ao risco, escolhendo grupos de nós sob risco que possuam intenso comércio entre si. Seria possível, na ocasião de erradicação de uma doença específica, dividir um programa sanitário em etapas nas quais um grupo de municípios é sanitizado por vez, caso esses municípios apresentem baixo risco de reinfecção devido a movimentos oriundos de fora do grupo. Propriedades que compartilham preferências comerciais podem compartilhar práticas de manejo e biossegurança, e serem selecionadas para um programa sanitário direcionado. A Organização Mundial de Saúde Animal (OIE), define em seu Terrestrial Animal Health Code \citep{OIE2010}, os conceitos de zoneamento, regionaliza\c{c}\~{a}o e compartimento. Dada a dificuldade de estabelecer e manter uma condição livre de doença para o território de um país inteiro, pode ser vantajoso estabelecer e manter uma subpopulação com condição sanitária diferenciada dentro de seu território. Assim, a delimitação de zonas sanitárias dentro de uma região pode levar em consideração não apenas municípios com as mesmas condições sanitárias, mas também regiões que possuam intenso comércio com estes. A análise de comunidades pode ajudar no planejamento da fronteira deste tipo de área, caso haja isolamento de uma subpopulação (p.ex, com restrição de movimentação), bloqueando o comércio entre fazendas de áreas diferentes: essa abordagem poderia otimizar a divisão de tal região e avaliar quanta perda econômica é esperada devido a restrição de comércio após uma zona específica ser estabelecida.\par Esses são apenas alguns exemplos. Como uma técnica exploratória, suas aplicações são virtualmente infinitas, facilitando a formulação de novas hipóteses, a avaliação de premissas, a seleção de futuras ferramentas analíticas e o planejamento de futuras coletas de dados.
\section{Definição de Comunidade e Circuitos Pecuários}
A tarefa de dividir uma rede em comunidades, intuitiva a princípio, não é realmente bem definida. Os principais elementos do problema, i.e. os conceitos de comunidade e cluster, não são rigorosamente definidos, requerendo um grau de arbitrariedade e/ou bom senso. Uma grande variedade de algoritmos e ferramentas foram desenvolvidos para revelar a estrutura comunitária de uma rede \citep{Fortunato2010}, e a escolha de qual algoritmo utilizar deve levar em consideração as definições e premissas do modelo, e se os mesmos são adequados para resolver o problema em questão.
\par Como notado por \citet{Bigras-Poulin2006}, em redes de movimentação de gados de corte as fazendas agem no geral como produtoras (algumas com grau de entrada igual a zero, chamadas de ``fazendas de cria") de gado, e abatedouros agem como escoadouros. Esses dois tipos de comportamento impõem um fluxo direcionado à rede. É portanto razoável assumir que a direção dos movimentos não pode ser ignorada sem a perda de informação importante, e é melhor usar ferramentas que levem em consideração essa direção sempre que possível.
\par As poucas aplicações recentes de detecção de comunidades em redes de trânsito animal que consideram a direção dos movimentos \citep{Green2009a,Green2011} usam a definição de comunidade dada por \citet{Newman2004c} e adaptada a redes direcionadas por \citet{Leicht2008b}: um conjunto de nós é uma comunidade se o número de ligações internas a este exceder o esperado para um mesmo conjunto de nós no modelo nulo \citep{Fortunato2010}. O que equivale a dizer que os nós que formam uma comunidade são mais densamente conectados entre si do que um conjunto similar que não forma uma comunidade. Essa definição é bastante útil e intuitiva, mas não é a única definição de comunidade possível.
\par
\begin{figure}[h]
\centering
\caption[Exemplo de circuito pecuário]{Exemplo de circuito pecuário}
\includegraphics[scale=0.2]{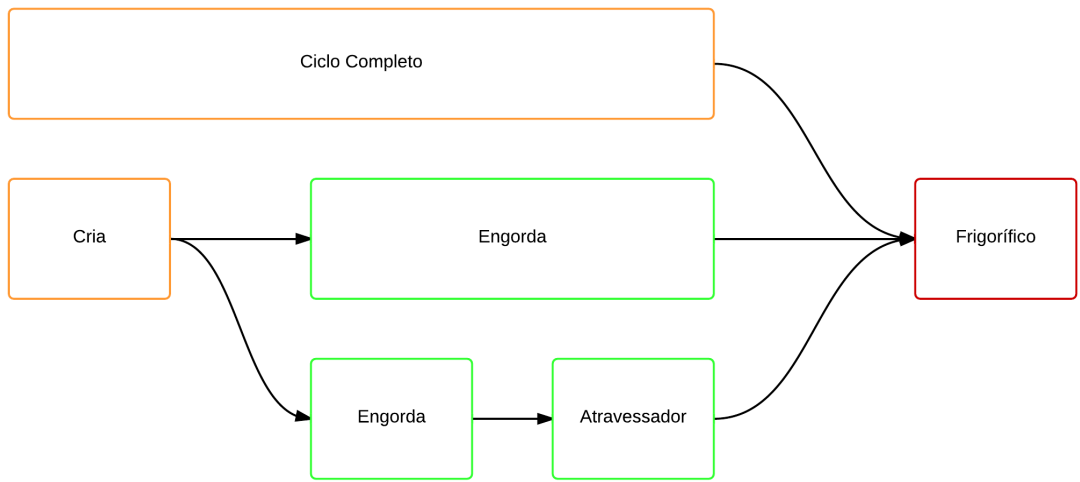}
\label{fig:circuito}
\end{figure}

 \citet{Kim2010} desenvolveram um método para detecção de comunidades que leva em consideração a direcionalidade e o peso das ligações, e que possui melhor desempenho que o método desenvolvido por~\citet{Leicht2008b} em alguns casos, trazendo uma definição de comunidade mais adequada em algumas situações.
Esse método é baseado no cálculo do LinkRank, um conceito derivado do PageRank do Google \citep{Page1999a}. Nessa definição, uma comunidade é definida como um grupo de nós nos quais um passeador aleatório permanece mais provavelmente do que o esperado pelo acaso. Se definirmos um circuito pecuário como o conjunto de localidades pelas quais um animal passará durante o ciclo produtivo (figura~\ref{fig:circuito}), então essa definição de comunidade é mais adequada para encontrar tais grupos, especialmente se desejamos compreender os fluxos entre zonas de produção de uma região ou país. Uma comunidade seria então definida como um grupo de locais onde um animal permanecerá mais provavelmente durante sua vida (o ciclo produtivo) do que o esperado pelo acaso.\par
Assim, o objetivo deste estudo foi implementar e testar um método de detecção de comunidades que identifique circuitos pecuários em uma rede de movimentação de animais.

\chapter*{2 MATERIAIS E MÉTODOS}
\addcontentsline{toc}{chapter}{2 MATERIAIS E MÉTODOS}\stepcounter{chapter}
\section{Base de dados}
Foi analisada a rede interna de movimentação de bovinos do Estado do Mato Grosso do ano de 2007 (movimentos realizados entre 01/01/2007 e 31/12/2007), fornecida pelo INDEA (Instituto de Defesa Agropecuária do Estado do Mato Grosso). Essa base de dados contém informações sobre 87.899 propriedades (fazendas, frigoríficos, eventos, etc) e 521.431 movimentos ao longo do ano, totalizando 15.844.779 de animais movimentados. Essa base foi analisada previamente por \citet{Negreiros2010}, que fez uma extensa busca para eliminar inconsistências cadastrais. O Estado do Mato Grosso é o terceiro maior do Brasil, com uma área de 903.357 km$^{2}$ e o maior rebanho nacional, com aproximadamente 28.757.438 bovinos (13\% do rebanho nacional em 2010), sendo em sua maioria rebanhos de exploração de corte \citep{InstitutoBrasileirodeGeografiaeEstatistica2010}. Está localizado na região centro oeste do país e possui 141 municípios.
\par
\begin{figure}[h]
\centering
\caption[Rede de movimentação de animais do Mato Grosso]{Rede de movimentação interna de animais do Mato Grosso durante o ano de 2007. Movimentos agrupados por municípios}
\includegraphics[scale=0.4]{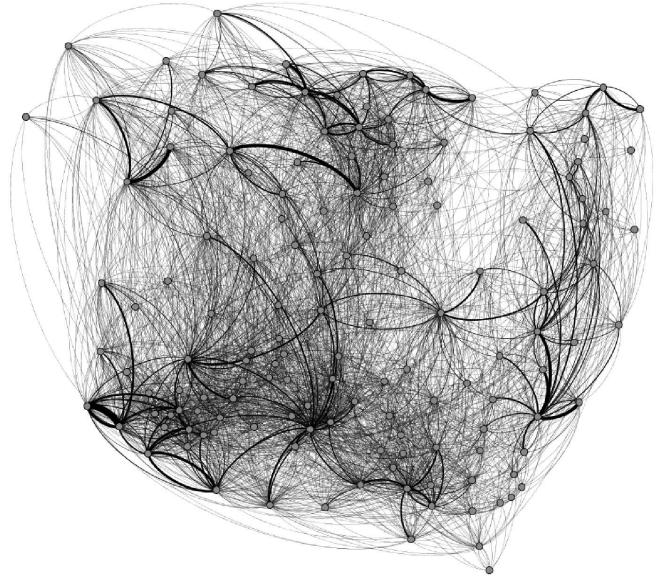}
\label{fig:redemt}
\end{figure}

 Foram extraídas informações sobre as propriedades de origem e destino e sobre o número de animais movimentados. Essas informações foram então resumidas para a menor divisão pública administrativa, agrupando os movimentos feitos por propriedades que pertenciam ao mesmo município. A rede analisada possui portanto 141 nós (municípios) e 3.980 arestas, estas últimas ponderadas pelo total de animais movimentados entre dois municípios durante o ano de 2007. A figura~\ref{fig:redemt} traz uma representação dessa rede, onde cada nó está localizado no centróide no respectivo município (aproximadamente).
\section{Detecção de circuitos pecuários}
Uma partição $C$ é uma divisão de um conjunto de $N$ objetos em clusters (ou grupos) não vazios e não sobrepostos. Os termos clusters, grupos e comunidades serão utilizados como sinônimos no presente trabalho. É possível representar uma partição $C$ como um vetor de tamanho $N$, no qual o i-ésimo elemento de $C$ é igual o grupo ao qual pertence o objeto $i$. Assim, uma partição $C = \{\{1\},\{1\},\{2\},\{2\}\}$ é a divisão de um conjunto de 4 objetos em dois grupos, nos quais os dois primeiros objetos foram designados ao primeiro grupo e os dois últimos objetos foram designados ao segundo grupo.
Da definição de comunidades emerge uma função chamada modularidade, a qual reflete a qualidade de qualquer partição em descrever a estrutura modular presente em uma rede. Quanto maior o valor de modularidade de uma partição, melhor essa partição descreve a estrutura da rede.
A função de modularidade proposta por \citet{Kim2010} pode ser calculada usando a seguinte equação:
\begin{equation} \label{eq:qlr}
Q^{lr} = \sum_{ij} [L_{ij} - \pi_i \pi_j] \delta_{g_i g_j} ,
\end{equation}
onde $Q^{lr}$ é a modularidade de uma dada partição; $L_{ij}$ é o valor de LinkRank da aresta que conecta os nós $i$ e $j$; $\pi_i$ é o valor de PageRank do nó $i$; $\pi_j$ é o valor de PageRank do nó $j$; e $\delta_{g_i g_j} = 1$ caso os nós $i$ e $j$ sejam designados à mesma comunidade e $\delta_{g_i g_j} = 0$ caso contrário. O valor do LinkRank de uma dada aresta entre os nós $i$ e $j$ é uma medida de sua importância, dada por:
\begin{equation} \label{eq:lr}
L_{ij} = \pi_i G_{ij} ,
\end{equation}
onde $\pi_i$ é o valor de PageRank \citep{Page1999a} do nó $i$ e $G$ é a Matriz Google (``Google Matrix", no original), uma matriz de probabilidade para o processo de passeio aleatório. A equação~\ref{eq:qlr} pode então ser formulada da seguinte maneira:
\begin{equation} \label{eq:gqlr}
Q^{lr} = \sum_{ij} \pi_i * [G_{ij} - \pi_j] \delta_{g_i g_j} .
\end{equation}
\par
\begin{figure}[h]
\centering
Figura 2.2 - PageRank calculado em rede simples\\
\includegraphics[scale=0.4]{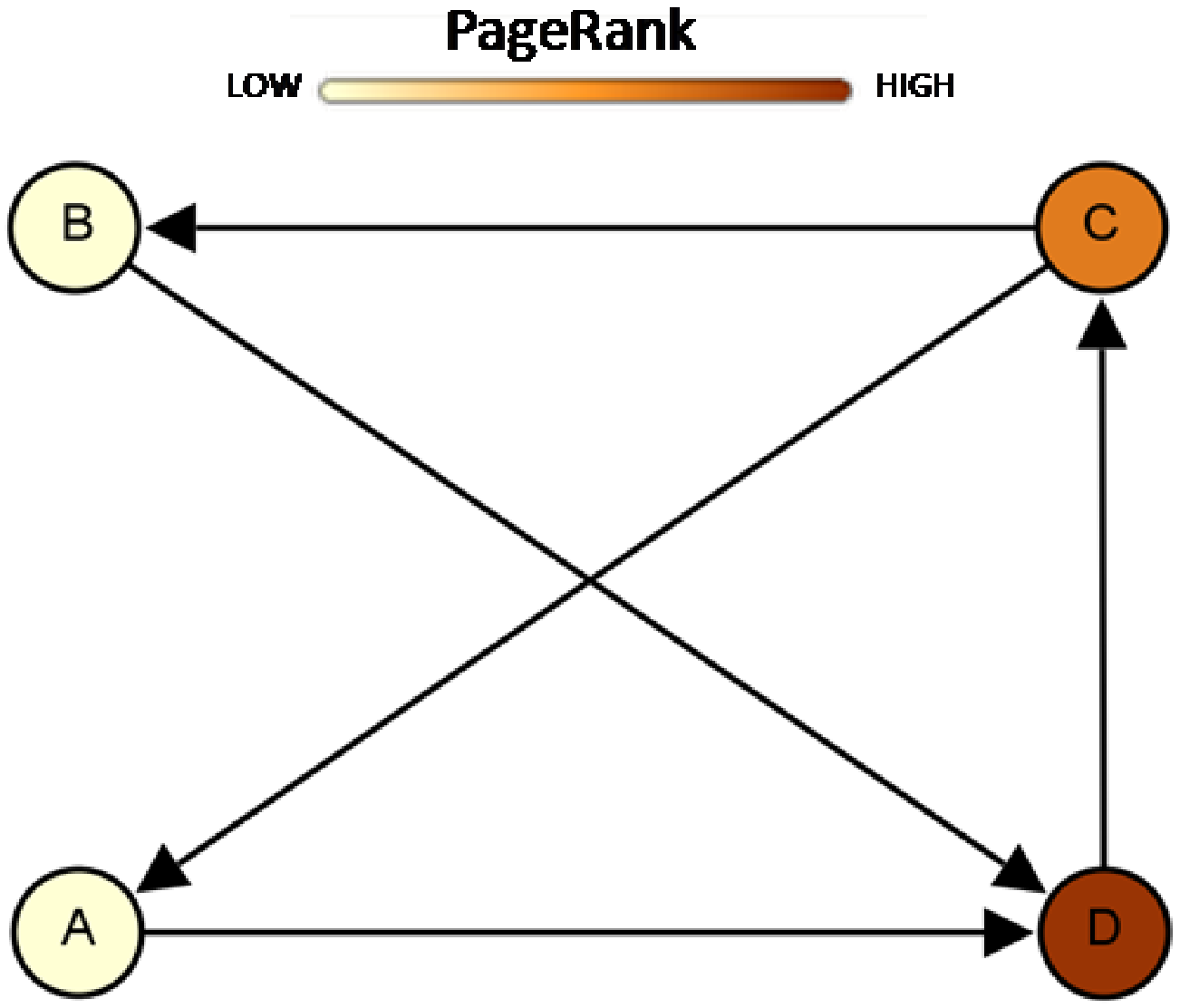}
\caption[PageRank calculado em rede simples]{Legenda:\\
Os nós A, B e C recebem o mesmo número de ligações (i.e. grau de entrada igual a 1). No entanto, o nó C possui um maior valor de PageRank que A e B, porque a ligação que C recebe provém de um nó importante (com maior grau de entrada, nó D), elevando assim seu PageRank. \label{fig:pagerankex} }
\end{figure}

 O PageRank é um algoritmo de análise de ligações que produz um ranking global da importância de todos os nós de uma rede, baseado na matriz de probabilidade do processo de passeio aleatório (também chamado de ``caminhada do bêbado"). Tradicionalmente, o cálculo do PageRank leva em consideração o grau de entrada de um nó e o grau de entrada de seus vizinhos. Alternativamente, o PageRank pode ser calculado levando em consideração o grau ponderado de entrada de um nó e o grau ponderado de entrada de seus vizinhos, o que foi feito neste trabalho, dada a importância do número absoluto de animais movimentados para a definição de circuito pecuário. O PageRank é uma medida propagada através da rede, baseada na premissa de que nós importantes recebem ligações de nós importantes (figura~\ref{fig:pagerankex}). Para maiores detalhes, veja \citet{AmyN.Langville2006} e \citet{Kim2010}.
\subsection{Número de Bell, máximos locais e máximos globais}
\par Como pode ser visto na equação \ref{eq:qlr}, a modularidade depende da alocação de $N$ nós em $0<g \le N$ comunidades não-vazias. Para encontrar a partição de maior modularidade, seria preciso calcular a modularidade de todas as partições possíveis. O número de partições possíveis para um conjunto de $N$ objetos é dado pelo número de Bell:
\begin{equation} \label{eq:bell}
B_N = \sum_{n=0}^{N-1} B_n \dbinom{N-1}{n} .
\end{equation}
\par Utilizando a equação \ref{eq:bell}, conclui-se que o número de partições possíveis para um conjunto de 141 objetos, é de:
28.\-911.\-910.\-353.\-139.\-891.\-247.\-190.\-947.\-943.\-924.\-637.\-289.\-025.\-637.\-429.\-885.\-820.\-537.\-824.\-520.\-005.\-904\-648.\-590.\-799.\-334.\-719.\-349.\-750.\-642.\-778\-141.\-473.\-707.\-684.\-781.\-558.\-911.\-314.\-718.\-429.\-702.\-080.\-529.\-710.\-551.\-046.\-049.\-912.\-485.\-833.\-910.\-034.\-612.\-983.\-281.\-492.\-385.\-095, ou aproximadamente $2.89 * 10^{178}$.
Considerando que o cálculo da modularidade de cada partição leve em média 5 milissegundos, o que parece ser uma boa estimativa, demoraria cerca de $1.6 * 10^{168}$ anos para calcular a modularidade de todas as partições possíveis. Se fossem utilizados 1.000 computadores apenas para essa tarefa, o tempo seria reduzido para $1.6 * 10^{165}$ anos. Se o cálculo fosse otimizado para ser realizado a cada ciclo do processador de um computador, utilizando um processador moderno capaz de realizar 5 milhões de cálculos por segundo (5GHz), ainda assim seriam necessários $6.4 * 10^{160}$ anos, utilizando 1.000 computadores apenas para essa tarefa.
Se faz necessário, portanto, um método para descobrir partições com altos valores de modularidade sem ter de calcular a modularidade de todas as partições existentes.
\par Um método possível seria começar com uma partição qualquer, digamos $C = \{\{1\},\{2\},...,\{N\}\}$ e calcular a modularidade dessa partição. Então em cada passo poderíamos sortear um nó e verificar se a mudança desse nó para uma comunidade aleatória forma uma partição de maior modularidade. Caso a nova partição possua uma modularidade maior, passamos a utilizar essa nova partição. Propomos então uma nova mudança e continuamos até que não seja possível, com qualquer mudança, aumentar a modularidade.
\par
\begin{figure}[h]
\centering
Figura 2.3 - Exemplo de função com dois máximos locais e um global\\
\includegraphics[scale=0.5]{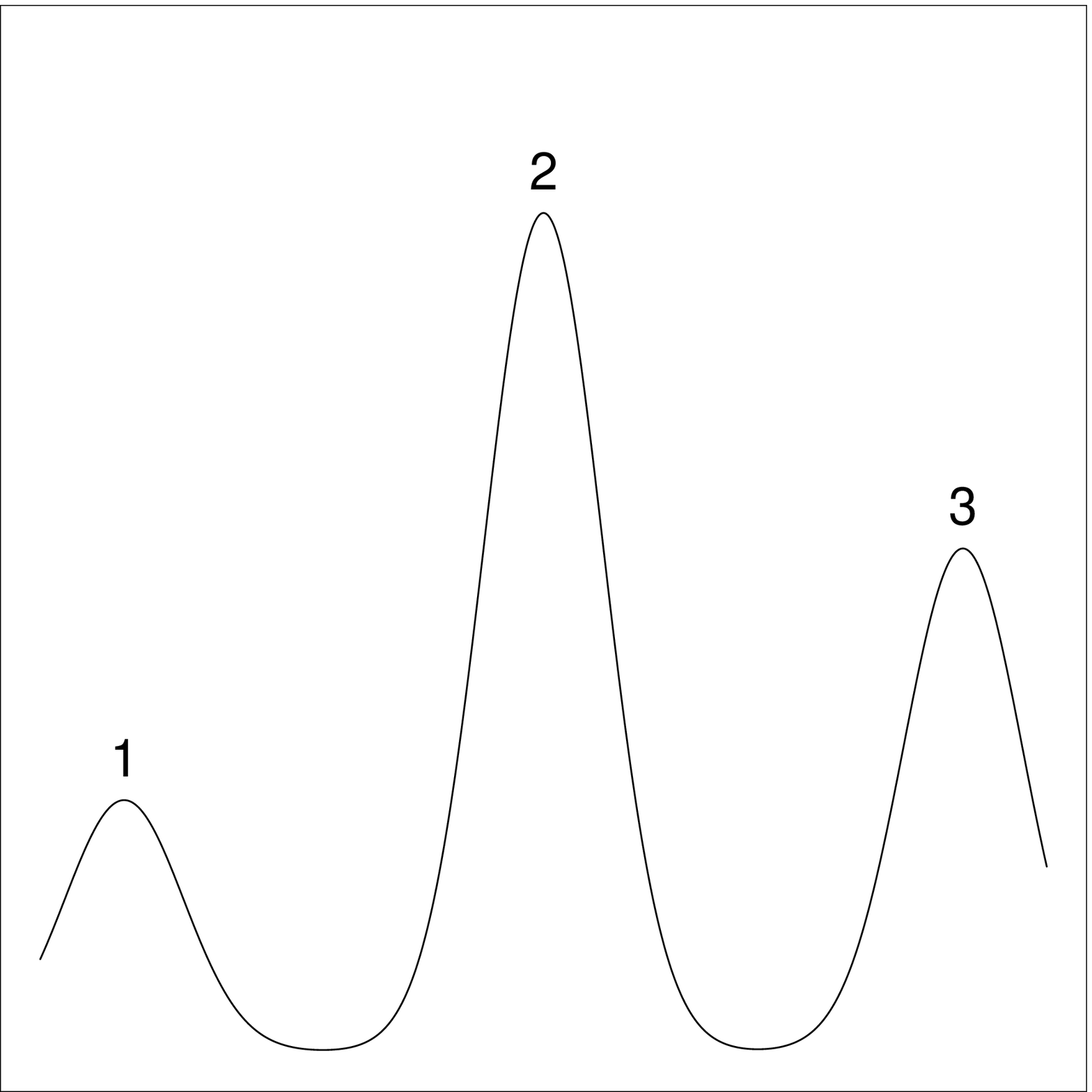}
\caption[Exemplo de função com dois máximos locais e um global]{Legenda:\\
Os pontos 1 e 3 indicam máximos locais. O ponto 2 indica o máximo global da função. \label{fig:maximos} }
\end{figure}

 A desvantagem desse método é que existe a possibilidade de obtermos uma partição de baixa modularidade, mas que a partir dela nenhuma mudança única resulte em aumento de modularidade. Essa situação é conhecida como ``máximo local" (figura~\ref{fig:maximos}). O máximo de uma função é chamado de ``máximo global". É necessário, portanto, um método capaz de encontrar o máximo global de uma função sem que seja necessário calcular a modularidade de todas as partições possíveis do conjunto de dados.
\subsection{Otimização da função de modularidade}
Para otimizar a função de modularidade e encontrar uma partição ótima da rede, foi utilizado o algoritmo de Arrefecimento Simulado (Simulated Annealing - SA) \citep{Kirkpatrick1983a}. Esse algoritmo é um método de otimização baseado em um procedimento probabilístico que evita que a solução encontrada pertença a um máximo local, e já foi aplicado para a otimização de modularidade anteriormente \citep{Guimera2005,Kim2010,Good2010}.
\par No estado inicial, cada n\'{o} \'{e} designado \`{a} sua própria comunidade. Em cada passo do algoritmo, um nó aleatório é designado \`{a} uma comunidade aleatória, e a mudança $\Delta Q^{lr}$ na modularidade é computada:
\begin{equation} \label{eq:dqlr}
\Delta Q^{lr} = Q^{lr}_n - Q^{lr}_o ,
\end{equation}
onde $Q^{lr}_n$ é a modularidade da nova partição e $Q^{lr}_o$ é a modularidade da partição anterior. Se $\Delta Q^{lr} > 0$, a mudança é aceita e a nova partição é assimilada para o próximo passo. Se $\Delta Q^{lr} \le 0$ então há uma probabilidade $P(\Delta Q^{lr})$ de que a mudança será aceita:
\begin{equation} \label{eq:pdqlr}
P(\Delta Q^{lr})=exp(\frac{\Delta Q^{lr}}{T}) ,
\end{equation}
onde $T$ \'{e} um n\'{u}mero real positivo, representando a temperatura atual do sistema. 
Talvez valha a pena mencionar que $T$ n\~{a}o se trata de uma temperatura real. O m\'{e}todo de arrefecimento simulado foi inspirado na termomecânica, e a repetição desse processo durante tempo suficiente simula o movimento de \'{a}tomos de um material sob processo de recozimento \`{a} uma temperatura $T$, utilizado para obtenção de estados de baixa energia de um sólido, da\'{i} o termo.
\par
\begin{figure}[p]
\centering
\caption[Esquema de funcionamento do algoritmo de Arrefecimento Simulado]{Esquema de funcionamento do algoritmo de Arrefecimento Simulado}
\includegraphics[scale=0.2]{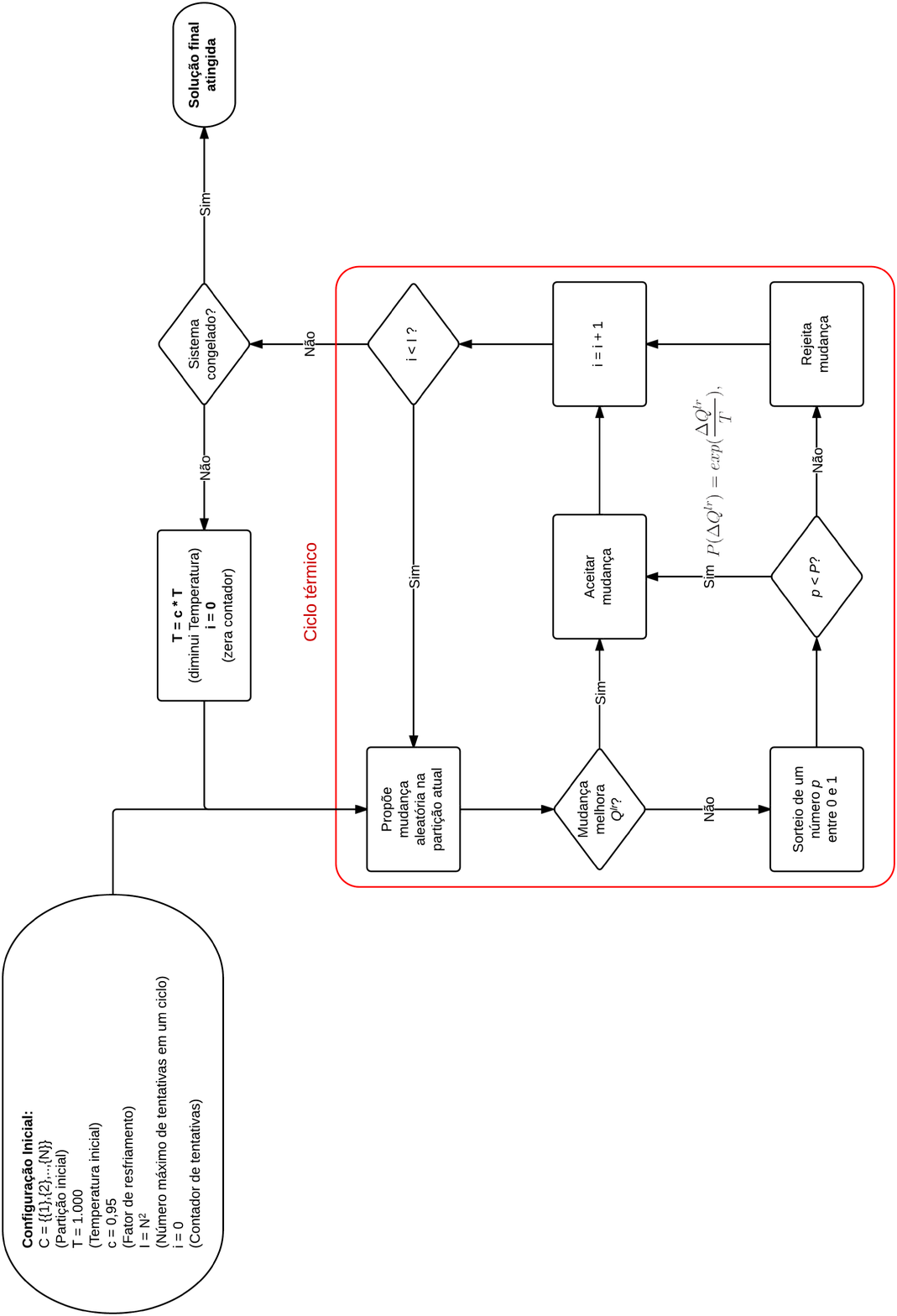}
\label{fig:SA2}
\end{figure}

 Depois de $I$ iterações			
o sistema \'{e} ``resfriado", multiplicando a temperatura presente $T$ por um fator de resfriamento $c$ ($0 < c < 1$). Este processo se repete até que o sistema esteja ``congelado", ou seja, nenhuma mudança proposta é aceita durante o ciclo de uma temperatura.
Isso significa que em altas temperaturas o sistema ir\'{a} explorar muitas configurações, e à medida em que o algoritmo avança e a temperatura decresce, o sistema passa a se concentrar nas melhores soluções, evitando mudanças que decresçam muito o valor de modularidade. Um esquema simplificado se encontra na figura~\ref{fig:SA2}
\par O número limite de iterações $I$, o valor da temperatura inicial $T_i$ e o fator de resfriamento $c$ devem ser ajustados para cada função específica a ser otimizada. Neste trabalho foram inicialmente utilizados os valores sugeridos por \citet{Guimera2005}, posteriormente ajustados até a obtenção de um desempenho satisfatório, de maneira que o algoritmo teste um conjunto suficientemente grande de soluções. Os valores foram ajustados em $I = N^2$ ($N$ = número de nós da rede), $T_i = 1.000$ e $c = 0,95$.
Para reduzir o tempo gasto pelo algoritmo, foi gerada uma matriz de Modularidade $Mm$, de tamanho $N$ x $N$ onde $Mm_{ij}$ é o resultado de $L_{ij} - \pi_i\pi_j$, o termo dentro dos colchetes da equação~\ref{eq:qlr}. Assim, essa matriz $Mm$ contém a contribuição à modularidade total caso os nós $i$ e $j$ sejam designados à mesma comunidade, possibilitando o cálculo direto de $\Delta Q^{lr}$, evitando o cálculo da modularidade total da partição testada a cada passo.
\subsection{Obtenção de uma solução única}
\citet{Good2010} mostraram que redes com estrutura modular apresentam frequentemente um alto número de degenerações, partições de alta modularidade que escondem a partição situada no máximo global (melhor solução). Assim, frequentemente redes exibirão não apenas um claro máximo global de modularidade, mas um platô de soluções, todas com alta modularidade, não muito diferentes entre si. Embora isso pareça impor um obstáculo para a detecção de comunidades dentro de uma rede, na verdade mostra a possibilidade de escolher uma divisão da rede que atenda necessidades específicas de uma situação sem violar a definição de comunidade previamente definida. Ao invés de analisar todas as partições de alta modularidade, pode ser útil extrair padrões comuns dessas partições, revelando onde elas concordam, i.e. quais nós aparecem juntos em todas as soluções encontradas. A solução final pode ser encarada como ``blocos de construção" (``building blocks"), os quais podem ser submetidos posteriormente a um método de aglomeração adequado.
\par Para resumir a informação obtida, foram selecionados os melhores resultados (partições com valores de modularidade até 1\% abaixo do máximo encontrado) e identificados como comunidades os grupos de municípios que foram designados para o mesmo cluster em todas as partições selecionadas. Essa solução foi obtida pela multiplicação elemento a elemento (produto Hadamard) das matrizes de similaridade de cada partição selecionada:
\begin{equation} \label{eq:cores}
Sm = Sm_1 \circ Sm_2 \circ ... \circ Sm_n ,
\end{equation}
onde $Sm$ é a matriz de similaridade correspondente à partição final $C$ e $Sm_n$ é a matriz de similaridade correspondente à partição $C_n$. Uma matriz de similaridade é uma matriz esparsa de tamanho $N$ x $N$, onde o elemento $Sm_{ij} = 1$ se os nós $i$ e $j$ pertencem ao mesmo cluster e $0$ caso contrário. Tendo como estrutura comunitária a partição $C$, foi possível analisar o comércio de animais intra e intercomunitário.
\section{Avaliação do desempenho do algoritmo}
\begin{figure}[h]
\centering
\caption[Representação gráfica da Variação de Informação]{Representação gráfica da Variação de Informação \citep{Meila2007a}}
\includegraphics[scale=0.2]{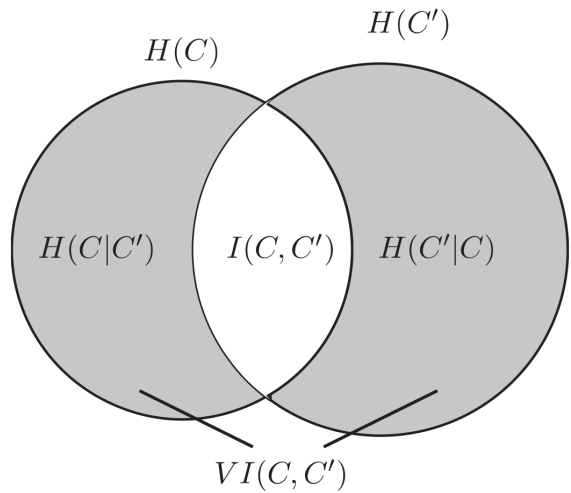}
\label{fig:vimeila}
\end{figure}

Uma maneira útil de avaliar se as partições obtidas são similares ou diferentes entre si é a distância de Variação de Informação (Variation of Information - VI) \citep{Meila2007a}. A distância VI é uma métrica baseada em teoria da informação e entropia, e mede a quantidade de informação não compartilhada entre duas partições.
A entropia associada a uma partição $C$ é dada pela seguinte equação:
\begin{equation}
H(C) = - \sum_{g=1}^g P(g) \log P(g) ,
\end{equation}
onde $P(g)$ é a probabilidade de escolher um objeto aleatório da partição $C$ que pertença à comunidade $g$:
\begin{equation}
P(g) = \frac{n_g}{N} ,
\end{equation}
onde $n_g$ é o número de objetos pertencentes à comunidade $g$.
Logo $H(C)$ provê a incerteza total a respeito de $C$. Define-se $P(g,g')$ como a probabilidade de um objeto pertencer à comunidade $g$ na partição $C$ e simultaneamente à comunidade $g'$ na partição $C'$, ou seja, a distribuição conjunta das variáveis aleatórias associadas às duas partições:
\begin{equation}
P(g,g') = \frac{| C_{g} \cap C'_{g'} |}{N} .
\end{equation}
Então a informação mútua entre as partições $C$ e $C'$ é dada por:
\begin{equation}
I(C,C') = \sum_{g=1}^G \sum_{g'=1}^{G'} P(g,g') \log \frac {P(g,g')} {P(g)P'(g')} ,
\end{equation}
e essa é a quantidade de incerteza sobre $C'$ que é reduzida se conhecermos $C$. A Variação de Informação (figura~\ref{fig:vimeila}) pode então ser definida da seguinte maneira:
\begin{equation}
VI(C,C') = H(C) + H(C') - 2I(C,C') .
\end{equation}
\par Quando aplicada a duas partições iguais $VI=0$. Quando aplicada a partições extremas que não compartilham nenhuma informação, como por exemplo a partição onde cada nó pertence à sua própria comunidade $C = \{\{1\},\{2\},...,\{N\}\}$ e a partição onde todos os nós pertencem à mesma comunidade $C'=\{g\}$, o valor de VI é o máximo possível, $VI = \log N$ (onde $N$ é o tamanho das partições analisadas).
Essa distância foi calculada entre a solução final, todas as soluções dadas pelo algoritmo SA e 50 partições aleatórias.
Esse conjunto de distâncias pode ser organizado em uma matriz de distância simétrica $D$, onde $D_{ij} = D_{ji}$, que representa a VI entre as partições $C_i$ e $C_j$.
\subsection{Escalonamento multidimensional}
Para facilitar a interpretação da magnitude das distâncias encontradas, optou-se por projetar a matriz $D$ em um plano euclidiano utilizando a técnica de escalonamento multidimensional. O escalonamento multidimensional (Multidimensional Scaling - MDS) constrói uma configuração de $N$ pontos em um espaço euclidiano tal que as distâncias entre pontos coincida, tanto quanto possível, as dissimilaridades originais entre $N$ objetos \citep{Cox2000}. 
\section{Softwares utilizados}
A função de modularidade, o algoritmo SA e o código para obtenção de uma solução única foram implementados no software R \citep{R2010}. Os mapas foram produzidos no software R \citep{R2010}, utilizando os seguintes pacotes: ``PBSmapping'' \citep{Schnute2010}, ``maptools" \citep{Lewin-Koh2011}, ``sp" \citep{Bivand2008} e ``foreign" \citep{RDevelopmentCoreTeam2011}. As representações gráficas de redes foram produzidas no software Gephi  \citep{Bastian2009}. As distâncias VI  foram calculadas utilizando o pacote ``mcclust" \citep{Fritsch2009}.

\chapter*{3 RESULTADOS}
\begin{figure}[h]
\begin{center}
Figura 3.1 - Partições amostradas e superfície de modularidade.\\
\includegraphics[scale=0.5]{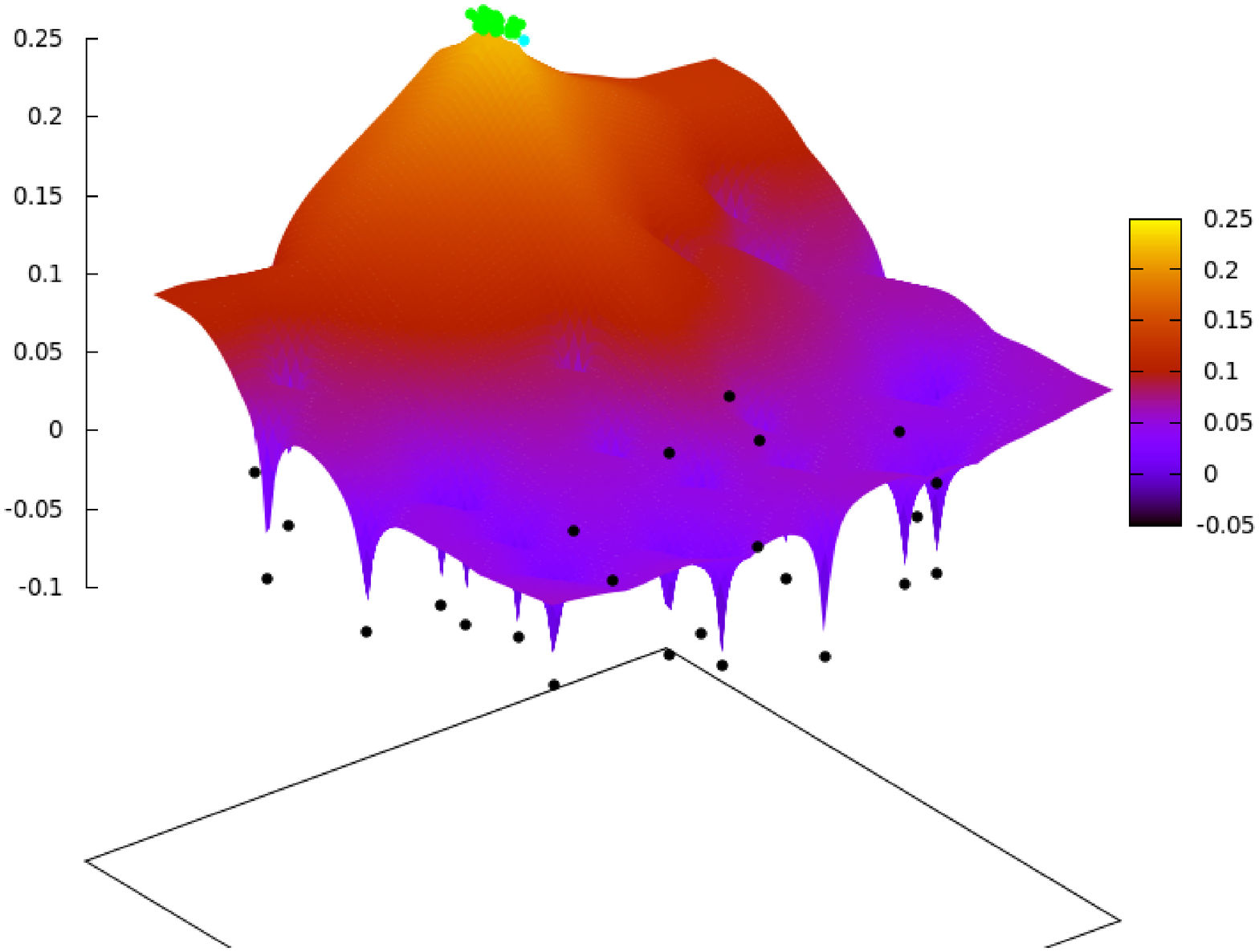}
\caption[Superfície de modularidade]{Legenda:\\
Distância entre os pontos nos eixos $x$ e $y$ representam a distância VI (stress = 23.89). Pontos verdes são partições obtidas pelo algoritmo SA, pontos pretos são partições aleatórias e o pronto azul claro é a partição final. A superfície colorida é a função de modularidade, reconstruída a partir de todas as partições analisadas. Partições obtidas pelo algoritmo SA estão localizadas no máximo da função de modularidade, muito próximas umas das outras quando comparadas com as partições aleatórias. Ainda, a partição final permanece perto das soluções do SA, e sua flexibilidade é obtida com o sacrifício de uma pequena parcela de sua modularidade.
\label{fig:surface}}
\end{center}
\end{figure}

\begin{figure}[h]
\centering
\caption[Partições obtidas com Arrefecimento Simulado e superfície de modularidade]{Partições obtidas com Arrefecimento Simulado e superfície de modularidade}
\includegraphics[scale=0.5]{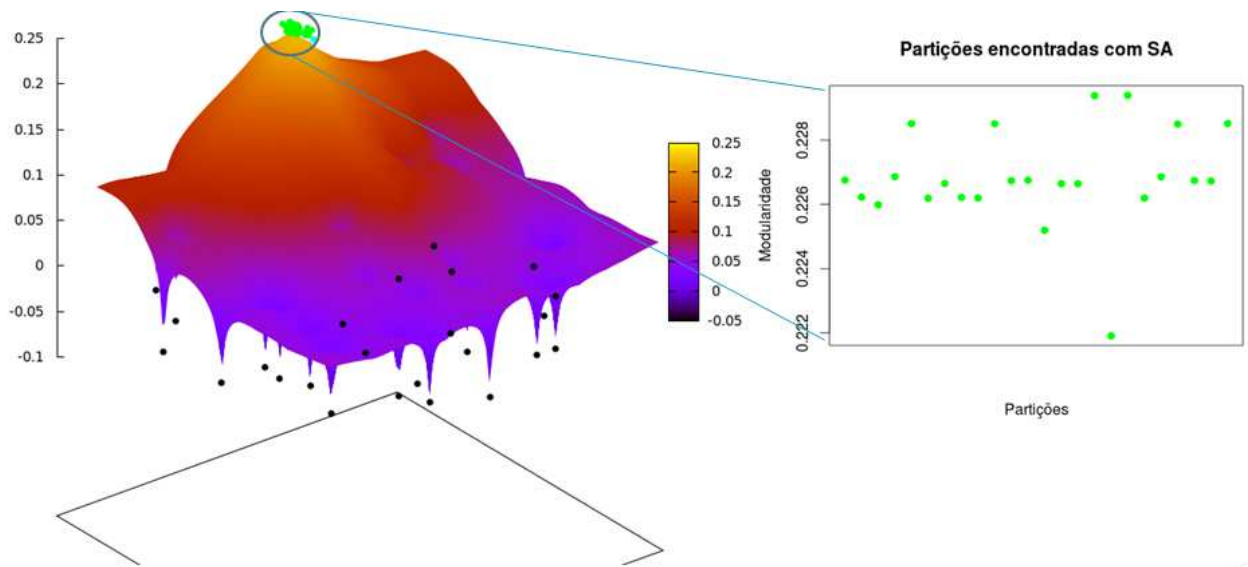}
\label{fig:supmods}
\end{figure}

\addcontentsline{toc}{chapter}{3 RESULTADOS}\stepcounter{chapter}
Após 50 iterações, o algoritmo SA foi capaz de identificar 24 partições únicas que maximizam a função de modularidade. Essas partições são muito similares entre si, tanto em termos de informação como em valores de modularidade (figura~\ref{fig:surface} e figura~\ref{fig:supmods}). 
\par
\begin{figure}[h]
\begin{center}
Figura 3.3 - Circuitos pecuários do Mato Grosso (detalhado)\\
\subfigure[Movimentos internos]{
\includegraphics[scale=0.16]{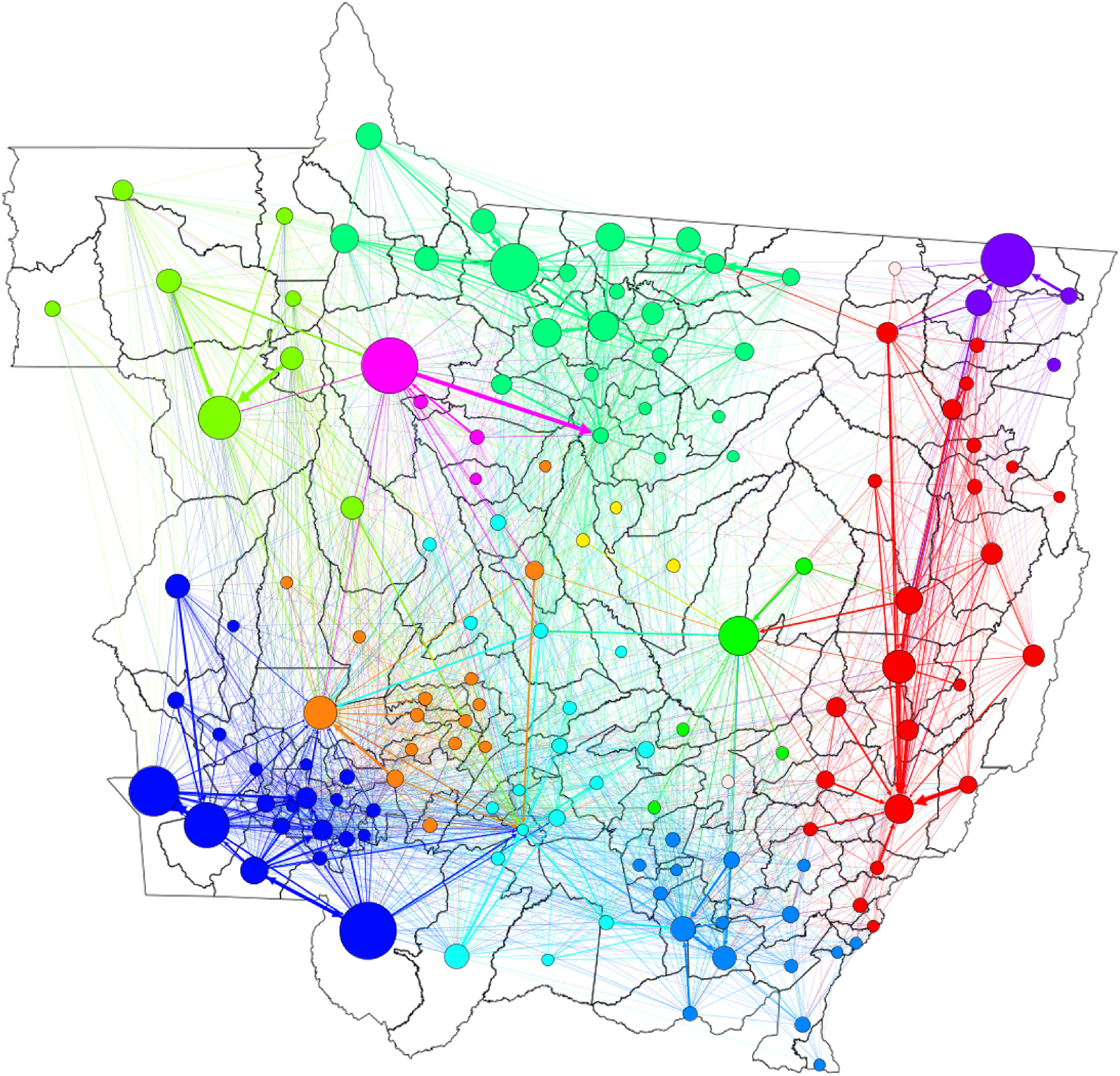}
\label{fig:loop}}
\subfigure[PageRank]{
\includegraphics[scale=0.16]{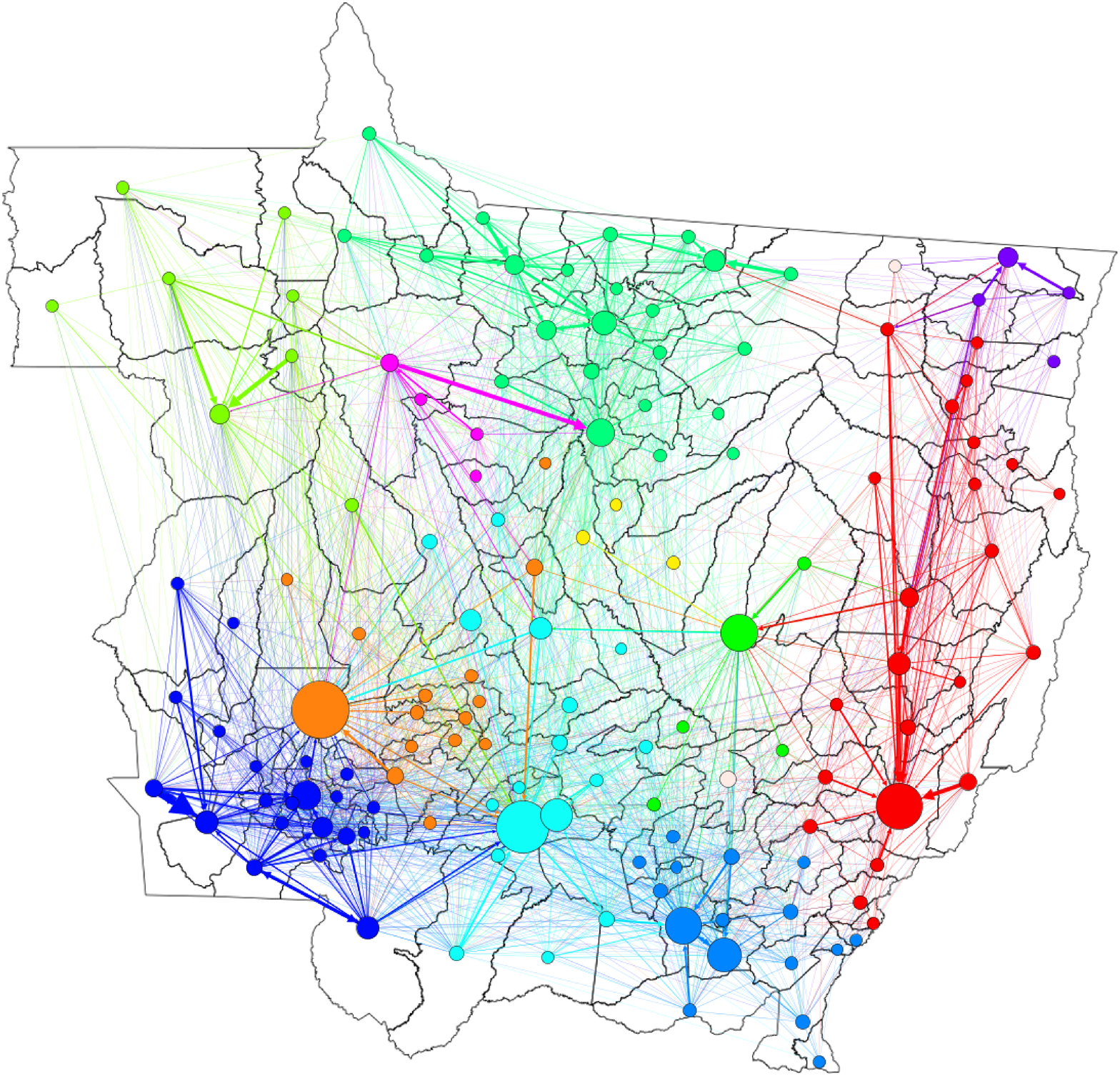}
\label{fig:pagerank}}
\caption[Mapa de comunidades]{Legenda:\\
A posição dos nós é dada pelo centroide do município correspondente (aproximado). As cores dos nós identificam as 11 comunidades encontradas. Nós brancos permaneceram desagrupados. O tamanho da aresta é proporcional ao número de animais movimentados entre dois municípios, e sua cor identifica a comunidade de origem de uma ligação. Na figura~\ref{fig:loop} o tamanho dos nós é proporcional ao número de animais comercializados internamente. Na figura~\ref{fig:pagerank} o tamanho dos nós é proporcional ao valor de PageRank.}
\label{fig:communitynetwork}
\end{center}
\end{figure}

\begin{figure}[t]
\begin{center}
Figura 3.4 - Circuitos pecuários do Mato Grosso.\\
\subfigure[Municípios]{
\includegraphics[scale=0.30]{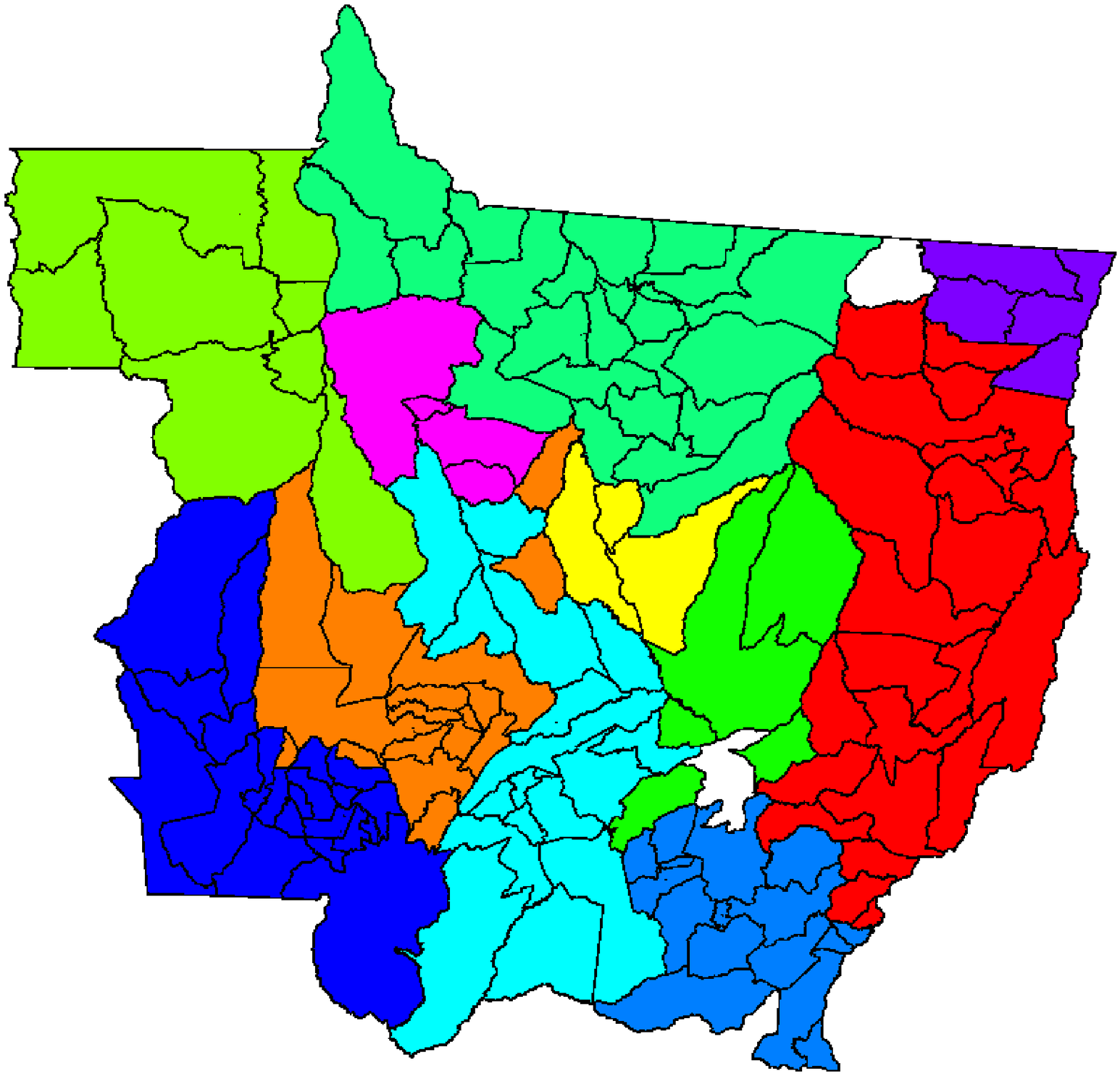}
\label{fig:counties}}
\subfigure[Fluxo de animais]{
\includegraphics[scale=0.15]{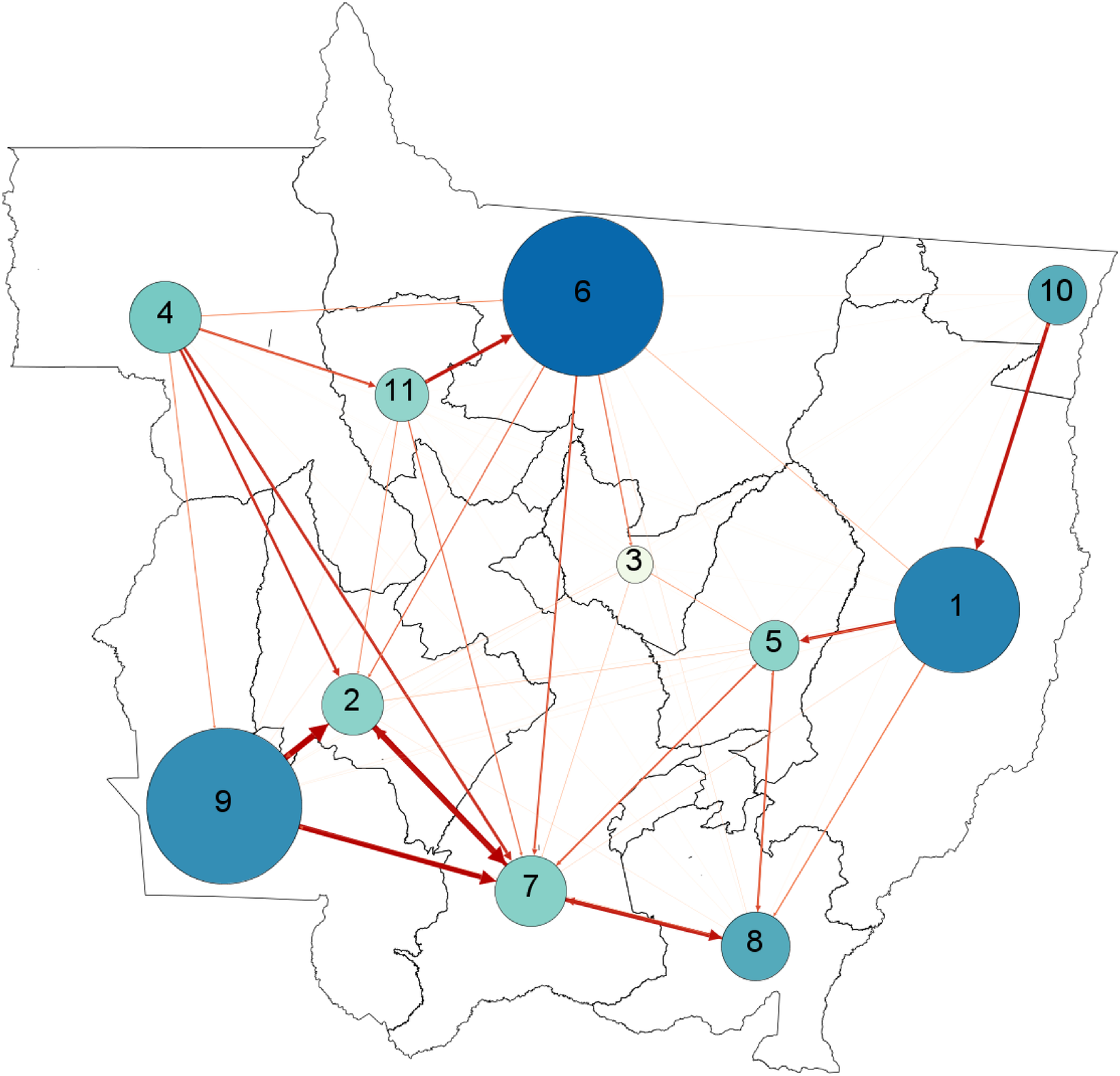}
\label{fig:flow}}
\caption[Mapa de comunidades]{Legenda:\\
A figura~\ref{fig:counties} mostra as 11 comunidades encontradas (municípios em branco permaneceram desagrupados). A figura \ref{fig:flow} mostra o comércio entre comunidades: números representam a identificação de cada comunidade (correspondendo às Tabelas \ref{outgoing} e \ref{ingoing}); a intensidade da cor e o tamanho das arestas estão correlacionados com o número de animais comercializados entre duas comunidades; o tamanho dos nós depende do número de animais transitados dentro de uma comunidade; e a intensidade de cor dos nós depende da probabilidade de que um animal vendido permaneça em sua própria comunidade (diagonal da Tabela~\ref{outgoing}). }
\label{fig:mapa}
\end{center}
\end{figure}

 A solução final, que resume a informação de todas as outras partições, retém um bom valor de modularidade, apresentando uma alta flexibilidade para a definição de clusters sem violar a definição de comunidade escolhida. Esta possui 11 comunidades, sendo que 2 municípios permaneceram desagrupados, dado que eles foram designados a comunidades distintas por duas ou mais partições (figuras ~\ref{fig:communitynetwork} e~\ref{fig:mapa}). Apenas três municípios foram designados a comunidades não adjacentes, mostrando que a solução em questão possui um claro padrão geográfico.
\par A partição encontrada mostra também um claro padrão comercial. Comunidades detectadas preferem comercializar animais internamente (Tabelas \ref{outgoing} e \ref{ingoing}). Em 10 das 11 comunidades encontradas um animal vendido têm mais chance de permanecer em sua própria comunidade (comprado por um nó interno à comunidade) do que ir para outra comunidade (Tabela~\ref{outgoing}). Embora a comunidade 3 mostre uma fraca preferência em comercializar animais internamente, devido em grande parte ao seu intenso comércio com a comunidade 5 e 6 (Tabelas~\ref{outgoing} e~\ref{ingoing}), o comércio interno é ainda maior do que o comércio desta com qualquer outra comunidade.
\begin{sidewaystable}[ht]
\begin{center}

\caption[Movimentos de saída (\%)]{Movimentos de saída (\%). $M_{ij} = \frac {W_{ij}} {kw_{i}^{out}}$  onde $W_{ij}$ é o número de animais movimentados da comunidade $i$ para a comunidade $j$, e $kw_{i}^{out}$ é o número de animais vendidos pela comunidade $i$ (grau ponderado de saída). Comunidades mostram clara preferência de vender animais internamente. O comércio realizado pelos 2 municípios não agrupados foi omitido.}
\scalebox{0.8}{
\begin{tabular}{rrrrrrrrrrrrr}
  \hline
 & 1 & 2 & 3 & 4 & 5 & 6 & 7 & 8 & 9 & 10 & 11 & Total de Animais \\ 
  \hline
1 & 91,60 & 0,02 & 0,00 & 0,00 & 3,68 & 0,69 & 0,27 & 1,44 & 0,02 & 1,56 & 0,00 & 2.519.752 \\ 
  2 & 0,02 & 71,31 & 0,13 & 0,54 & 1,35 & 0,80 & 19,21 & 0,17 & 6,26 & 0,00 & 0,08 & 924.083 \\ 
  3 & 0,00 & 10,54 & 39,23 & 0,00 & 23,73 & 10,69 & 14,72 & 0,04 & 1,05 & 0,00 & 0,00 & 89.107 \\ 
  4 & 0,00 & 6,53 & 0,00 & 76,02 & 0,01 & 2,17 & 7,04 & 0,02 & 2,41 & 0,00 & 5,80 & 1.228.075 \\ 
  5 & 5,56 & 0,30 & 0,73 & 0,00 & 70,91 & 0,27 & 8,78 & 10,57 & 0,42 & 0,00 & 0,01 & 523.644 \\ 
  6 & 0,08 & 1,11 & 1,07 & 0,19 & 0,00 & 94,70 & 1,74 & 0,18 & 0,14 & 0,00 & 0,74 & 3.371.829 \\ 
  7 & 0,28 & 9,44 & 0,14 & 0,17 & 3,23 & 1,11 & 72,46 & 9,73 & 2,77 & 0,00 & 0,37 & 1.265.583 \\ 
  8 & 2,14 & 0,42 & 0,00 & 0,02 & 3,39 & 0,20 & 8,17 & 83,66 & 0,78 & 0,01 & 0,01 & 1.016.299 \\ 
  9 & 0,02 & 5,31 & 0,01 & 0,22 & 0,02 & 0,07 & 4,50 & 0,16 & 89,60 & 0,01 & 0,07 & 3.439.689 \\ 
  10 & 15,18 & 0,00 & 0,02 & 0,00 & 0,22 & 0,33 & 0,00 & 0,03 & 0,01 & 82,57 & 0,01 & 724.604 \\ 
  11 & 0,00 & 4,10 & 0,05 & 3,26 & 0,00 & 16,66 & 5,52 & 0,01 & 0,57 & 0,00 & 69,83 & 656.241 \\ 
   \hline
\end{tabular}
}

\label{outgoing}
\end{center}
\end{sidewaystable}

\begin{sidewaystable}[ht]

\begin{center}
\caption[Movimentos de entrada (\%)]{Movimentos de entrada (\%). $M_{ij} = \frac{W_{ij}}{kw_{j}^{in}}$ onde $W_{ij}$ é o número de animais movimentados da comunidade $i$ para a comunidade $j$, e $kw_{j}^{in}$ é o número de animais comprados pela comunidade $j$ (grau ponderado de entrada). O comércio realizado pelos 2 municípios não agrupados foi omitido.}
\scalebox{0.8}{
\begin{tabular}{rrrrrrrrrrrr}
  \hline
 & 1 & 2 & 3 & 4 & 5 & 6 & 7 & 8 & 9 & 10 & 11 \\ 
  \hline
1 & 92,38 & 0,05 & 0,01 & 0,00 & 15,88 & 0,52 & 0,43 & 3,32 & 0,01 & 6,08 & 0,01 \\ 
  2 & 0,01 & 58,75 & 1,51 & 0,51 & 2,13 & 0,22 & 11,19 & 0,14 & 1,79 & 0,00 & 0,12 \\ 
  3 & 0,00 & 0,84 & 44,54 & 0,00 & 3,62 & 0,28 & 0,83 & 0,00 & 0,03 & 0,00 & 0,00 \\ 
  4 & 0,00 & 7,15 & 0,03 & 95,60 & 0,02 & 0,78 & 5,45 & 0,02 & 0,92 & 0,00 & 12,67 \\ 
  5 & 1,16 & 0,14 & 4,89 & 0,00 & 63,59 & 0,04 & 2,90 & 5,05 & 0,07 & 0,00 & 0,01 \\ 
  6 & 0,10 & 3,34 & 45,79 & 0,67 & 0,03 & 94,19 & 3,71 & 0,54 & 0,14 & 0,00 & 4,41 \\ 
  7 & 0,14 & 10,65 & 2,18 & 0,22 & 6,99 & 0,42 & 57,81 & 11,23 & 1,09 & 0,00 & 0,82 \\ 
  8 & 0,87 & 0,38 & 0,04 & 0,02 & 5,90 & 0,06 & 5,23 & 77,57 & 0,25 & 0,01 & 0,02 \\ 
  9 & 0,03 & 16,30 & 0,42 & 0,78 & 0,14 & 0,08 & 9,76 & 0,49 & 95,58 & 0,03 & 0,46 \\ 
  10 & 4,40 & 0,00 & 0,15 & 0,00 & 0,27 & 0,07 & 0,00 & 0,02 & 0,00 & 92,58 & 0,01 \\ 
  11 & 0,00 & 2,40 & 0,44 & 2,19 & 0,00 & 3,23 & 2,28 & 0,01 & 0,12 & 0,00 & 81,47 \\ 
  Total de Animais & 2.498.589 & 1.121.681 & 78.501 & 976.503 & 583.919 & 3.390.257 & 1.586.209 & 1.096.007 & 3.224.366 & 646.296 & 562.458 \\ 
   \hline
\end{tabular}
}

\label{ingoing}
\end{center}
\end{sidewaystable}

\chapter*{4 DISCUSSÃO}
\addcontentsline{toc}{chapter}{4 DISCUSSÃO}\stepcounter{chapter}

\par Ao enfrentar o desafio de definir de comunidades, a escolha de qual algoritmo utilizar deve levar em consideração as definições e premissas do modelo, e se os mesmos são adequados para resolver o problema em questão.
\par O método desenvolvido por \citet{Kim2010} é baseado no cálculo de LinkRank, uma medida baseada no PageRank e na matriz Google. O PageRank pode ser interpretado como a probabilidade de que um passeador aleatório encontre um determinado nó (utilizando uma taxa mínima fixa $\alpha$, chamada de probabilidade de teletransporte, para maiores detalhes consultar \citet{AmyN.Langville2006}). Já a matriz Google traz as probabilidades de caminhos de saída de um nó para um passeador aleatório que esteja percorrendo a rede. Portanto, quando aplicado a uma rede ponderada e direcionada de trânsito animal entre municípios, o método calcula os caminhos possíveis a serem percorridos por um animal aleatório, a quantidade de animais comercializados, a preferência comercial das propriedades de cada município, e a probabilidade de um animal aleatório transitar por um determinado município. Por ser uma medida propagada através da rede, nós que se encontram no fim do ciclo produtivo, como municípios contendo frigoríficos e fazendas de engorda, acumulam uma maior fração de PageRank.
\par Uma alteração no cálculo do Pagerank poderia ajustar de maneira mais precisa a distribuição de probabilidades referida, bastando para isso alterar a taxa $\alpha$ para um valor que reflita a reposição de animais no sistema como um todo. Ou, alternativamente, poder-se-ia alterar esse parâmetro para refletir a reposição de animais apenas em propriedades de cria e ciclo completo, por exemplo. Tais alterações não foram realizadas, visto não ser objetivo deste projeto avaliar esta probabilidade com tamanha precisão.
\par O método, portanto, deveria valorizar partições nas quais os animais possuam maior probabilidade de permanecer em uma comunidade do que sair desta. A única ressalva é que, como pode ser observado na equação~\ref{eq:qlr}, maior será a contribuição ao valor da modularidade da designação de dois nós a mesma comunidade quanto maior for a diferença entre o LinkRank ($L_{ij}$) e o produto de seus PageRanks ($\pi_i \pi_j$). Trabalha-se com a premissa, portanto, de que é esperado que nós de elevado PageRank estejam conectados uns aos outros. Essa característica traz importantes implicações. Em primeiro lugar, se dois nós $i$ e $j$ de elevado PageRank não possuírem um alto valor nas células correspondentes da matriz de passeio aleatório $G_{ij}$ e $G_{ji}$ (o que indica uma fraca conexão entre eles) esses nós provavelmente serão designados a comunidades diferentes, pois sua contribuição à $Q^{lr}$ será negativa (já que, de acordo com a equação~\ref{eq:gqlr}, se $G_{ij} < \pi_j$ então $Q^{lr}<0$). Com isso, nós que tendem a acumular grande parte do PageRank da rede, como municípios que possuam frigoríficos ou grandes comerciantes de gado, serão designados a comunidades diferentes a não ser que possuam elevado comércio entre si. Desta maneira, é mais provável que os municípios com os maiores valores de PageRank da rede formem comunidades que orbitem em volta deles, agregando municípios de valores mais baixos de PageRank mas com os quais possuam elevados valores nas células da matriz Google. 
\par Em segundo lugar, um par de municípios com baixo valor de PageRank será designado à mesma comunidade se possuir internamente uma fração representativa do seu comércio $G_{ij}$ e $G_{ji}$. Por último, caso um município de baixo valor de PageRank $i$ possua um comércio modesto com um município de alto valor de PageRank $j$ (representado por um valor médio em $G_{ij}$, mas de modo que $G_{ij} < \pi_j$) sua contribuição à $Q^{lr}$ será negativa. Obviamente, essas deduções atuam de maneira conjunta para os 141 municípios estudados, e o método de maximização utilizado busca a melhor combinação de municípios em grupos, de modo que se a contribuição de dois nós à modularidade for negativa ($Q^{lr}_{ij}~<~0$), mas superada pela agregação de ambos à mesma comunidade de um terceiro ($Q^{lr}_{im}~>>~0; Q^{lr}_{jm}~>>~0$), ambos serão designados ao mesmo grupo.
\par Dessa maneira, valoriza-se ligações fortes com nós de baixo PageRank, o que implica em descontar o efeito agregador que municípios de alto PageRank exercem na rede. Ainda, o método praticamente ignora o comércio feito de maneira ocasional (onde $G_{ij}$ é baixo), o que é desejável para uma técnica exploratória desta natureza, haja vista que o objetivo principal é realçar os grandes padrões de movimentação e reduzir a sobrecarga decorrente do excesso de informações causado por movimentos não-significativos.
\par Outro método possível de ser utilizado, projetado para redes direcionadas, é o método de \citet{Leicht2008b}, o qual utiliza a seguinte equação para o cálculo de modularidade:
\begin{equation} \label{eq:leicht}
Q = \frac{1}{m} \sum_{ij} [ A_{ij} - \frac{k_i^{in} k_j^{out}}{m}]\delta_{g_i,g_j} .
\end{equation}
Como pode ser visto na equação~\ref{eq:leicht}, o método de \citet{Leicht2008b} considera o número de ligações de cada nó, e pode ser adaptado para uma rede ponderada, alterando  a matriz de adjacência $A$ pela matriz contendo o peso das arestas $W$, e alterando os termos de grau $k_i^{in}$ e $k_j^{out}$ pelos graus ponderados $kw_i^{in}$ e $kw_j^{out}$. No entanto, o método não utiliza em seus cálculos a topologia da rede como um todo, apenas dados dos nós individualmente, como grau de entrada e grau de saída. Embora o método de \citet{Kim2010} também utilize características individuais de nós e arestas, como PageRank e LinkRank, esses são fruto de um cálculo que leva em consideração a topologia da rede estudada, sendo portanto influenciados por ligações indiretas a estes, como por exemplo o fluxo encadeado de animais entre vários municípios.
\par Embora a técnica de arrefecimento simulado tenha encontrado várias partições distintas, esses resultados mostram uma notória similaridade entre si, concordando na alocação em comunidades da maioria dos municípios, o que é evidenciado pela figura~\ref{fig:surface}, onde nota-se que as partições encontradas possuem baixa Variação de Informação entre si. É possível notar, na figura~\ref{fig:supmods}, que as partições obtidas também possuem valores muito similares de modularidade (entre 0,22 e 0,23).
\par Redes que possuem mais de uma partição com alto valor de modularidade são comuns e esperadas. Isso apresenta um obstáculo para escolher, dentre várias partições, qual seria a correta. No entanto, é possível perceber essa característica como uma vantagem, dado que possibilita a escolha da partição que melhor atenda às necessidades de um pesquisador ou tomador de decisão, mantendo a definição escolhida de comunidades. 
\par Extrair as concordâncias entre as diferentes partições é uma maneira útil de obter grupos pequenos que podem ser posteriormente aglomerados de maneira a atender os objetivos específicos de um estudo sem violar a definição de comunidade escolhida. 
\par Nota-se na figura~\ref{fig:surface} a baixa Variação de Informação entre a partição final, obtida através do produto Hadamard, e as partições encontradas pelo algoritmo de Arrefecimento Simulado. Isto evidencia que grande parte da informação contida nas 24 partições encontradas foi conservada após a operação que deu origem à solução final.
\par A figura~\ref{fig:counties} mostra que a partição obtida reflete um claro padrão comercial, no qual apenas três municípios foram designados a uma comunidade com a qual não possuem contato geográfico. Seria necessário verificar individualmente a razão deste fato curioso. Um deles, o município de Ipiranga do Norte, foi criado em 2005, desmembrado do município de Tapurah, que por sua vez foi designado a uma comunidade adjacente, mantendo o padrão encontrado no restante do estado. Uma possível explicação seria a de que, em 2007, o sistema do INDEA se encontrasse em fase de adequação à existência do município, atualizando dados cadastrais dos produtores da região.
\par Na figura~\ref{fig:pagerank} nota-se que muitas comunidades possuem um ou dois municípios com alto valor de PageRank. Na comunidade de cor laranja, no sudoeste do estado, o município de Tangará da Serra é o que se destaca com maior valor de PageRank. Na comunidade de cor azul clara, no centro sul do estado, os municípios são Cuiabá (a capital do estado) e Várzea Grande (situado na região metropolitana de Cuiabá). Na comunidade vermelha, na região leste do estado, o município de destaque é Barra do Garças. Na comunidade azul escura, no sudeste do estado, destacam-se os municípios de Rondonópolis e Pedra Preta. Na comunidade de cor verde clara, no centro leste do estado, nota-se o município de Paranatinga. Na comunidade verde escura, no centro norte do estado, destacam-se, de maneira mais modesta, os municípios de Sinop e Colider. Todos os municípios mencionados são conhecidos por conterem um ou mais frigoríficos em seus territórios, o que corrobora a hipótese de que nós no fim da cadeia produtiva concentram grande parte do PageRank da rede e de suas respectivas comunidades.
\par Nota-se, através das tabelas~\ref{outgoing} e~\ref{ingoing}, que essa solução final é consistente com a ideia de circuitos pecuários, visto que um animal vai mais provavelmente permanecer em sua comunidade de origem, ao invés de ser transferido para outra comunidade qualquer. A única exceção à essa regra é a comunidade 3. Como pode ser visualizado na figura~\ref{fig:mapa}, esta comunidade é formada por apenas três municípios da região central do estado, todos com baixo valor de PageRank (figura~\ref{fig:pagerank}), indicando, possivelmente, a ausência de frigoríficos na região. Na tabela~\ref{ingoing} vemos que esta é a comunidade que menos comprou animais (cerca de 80.000), e que a mesma adquire animais prioritariamente da comunidade 6 (45,79\% dos animais comprados), situada imediatamente ao norte desta. Na tabela~\ref{outgoing} vemos que esta é a comunidade que menos vende animais (cerca de 90.000), prioritariamente para as comunidades 2 (10,54\%), 5 (23,73\%), 6 (10,69\%) e 7 (14,72\%). À exceção da comunidade 6, todas essas se situam ao sul da comunidade 3 (figura~\ref{fig:flow}), caracterizando a mesma como uma ponte entre municípios do norte (comunidade 6) e municípios do sul (comunidades 2, 5 e 7).
\par Esse padrão de movimentação, caracterizado pelo fluxo de animais no sentido norte-sul, pode ser observado em outras regiões do estado, com o auxílio da figura~\ref{fig:flow} e da tabela~\ref{outgoing}. Começando pela região nordeste, por exemplo, vemos que a comunidade 10 envia 15,18\% de seus animais vendidos para a comunidade 1, que por sua vez envia 3,68\% de seus animais vendidos para a comunidade 5, que por sua vez envia animais para as comunidades 7 (8,78\%) e 8 (10,57\%). Já na região noroeste, nota-se a existência de dois fluxos distintos. Um no sentido oeste-leste, começando pela comunidade 4, que envia animais para a comunidade 11 (5,8\%), que por sua vez envia animais para a comunidade 6 (16,66\%). E outro no sentido norte-sul, começando também pela comunidade 4, enviando animais para as comunidades 2 (6,53\%) e 7 (7,04\%). Paralelamente, a comunidade 11 envia animais para as comunidades 2 (4,1\%) e 7 (5,52\%), e a comunidade 6 envia animais para as comunidades 2 (1,11\%), 3 (1,07\%) e 7 (1,74\%). Esse fluxo pode ser causado pela concentração da população humana na região sul do estado, ao redor da capital Cuiabá, ou pela proximidade desta região com o estado de São Paulo, onde localiza-se grande mercado consumidor de carnes.
\par Fica evidenciado que após uma análise da estrutura da indústria pecuária em um estado, é possível fornecer informações importantes para aplicações diversas na área de defesa sanitária, principalmente na vigilância e no controle de enfermidades. Ao obter uma partição na qual animais possuam a tendência de permanecer em suas regiões, um estudo de circulação viral pode levar em conta as diferentes comunidades encontradas na sua estratificação, buscando assim dividir a área estudada em diferentes zonas de produção. Considerando a relação entre os circuitos pecuários, seria possível, na ocasião de erradicação de uma doença específica, atuar sequencialmente nessas diferentes zonas de produção, começando por aquelas que possuam menor probabilidade de reinfecção, ou que dependam menos, do ponto de vista econômico, do ingresso de animais de outras regiões, na eventual possibilidade de criar zonas com diferentes status sanitário com restrição de comércio.
\par Uma possível análise seria agregar as informações dos fluxos de animais com a infraestrutura do estado para locomoção de animais e suas particularidades ambientais, como grandes estradas e rotas de trânsito que são bloqueadas na estação de chuvas, a fim de estudar como o fluxo de animais é influenciado por estes. Nota-se, por exemplo, na figura~\ref{fig:communitynetwork} que na região nordeste do estado há uma extensão no sentido norte-sul entre as comunidades 6, 5, 1 e 10 (figura~\ref{fig:flow}) com baixa densidade de ligações. Esta extensão corresponde à área do Parque Indígena do Xingú, criado em 1961, que possui aproximadamente 27.000 km$^2$.
\par A solução final apresenta um padrão geográfico e comercial claro, característica crucial para aplicações em medicina veterinária preventiva, além de possuir uma interpretação clara e útil em redes de comércio onde ligações emergem das escolhas de nós comerciantes, e esses últimos são fortemente influenciados pelos ``hubs" (nós com alto valor de PageRank), os quais normalmente apresentam alta competitividade quando comparados com outros nós da rede. A solução final também é consistente do ponto de vista de informação, dada a baixa Variação de Informação entre esta e as partições encontradas pelo Arrefecimento Simulado. A técnica apresentada permite extrair informações relevantes acerca do fluxo de uma rede de trânsito animal, permitindo um melhor entendimento da relação existente entre suas zonas de produção.

\chapter*{5 CONCLUSÕES}
\addcontentsline{toc}{chapter}{5 CONCLUSÕES}\stepcounter{chapter}
Vimos que o método descrito nessa tese pode revelar a estrutura comunitária de uma rede direcionada de trânsito animal. Acreditamos que essa abordagem é apropriada para redes de trânsito animal e circuitos pecuários, assim como outras redes onde o caminho percorrido por um passeador aleatório (ou as escolhas dos nós em estabelecer ligações com outros nós) está incluída na definição de comunidade.
Essa abordagem pode ser utilizada para melhor compreender o fluxo interno de uma rede, ajudar no planejamento de um sistema de vigilância baseado em risco, melhorar modelos preditivos, guiar uma amostra estratificada, determinar áreas alvos para programas sanitários e definir zonas para restrição de movimentação.
\end{doublespace}

\newpage
\addcontentsline{toc}{chapter}{REFERÊNCIAS}\stepcounter{chapter}
\bibliographystyle{teste}
\raggedright
\bibliography{library}

\pagestyle{myheadings}

\chapter*{APÊNDICE - Códigos desenvolvidos}
\addcontentsline{toc}{chapter}{APÊNDICE}\stepcounter{chapter}
\markboth{}{}
\doublespacing
\setlength{\parindent}{50pt}
\setlength{\parskip}{1.5ex}
\section*{Algoritmo}
\par O código deve ser utilizado na seguinte ordem:
\begin{enumerate}
\item Em uma rede com $N$ nós, criar uma matriz $W$ de tamanho $N$x$N$, onde $W_{ij}$ é a quantidade de animais que o nó $i$ enviou ao nó $j$;
\item Calcular a matriz de hiperlinks normalizada por linhas $H$;
\item Calcular o vetor de PageRank $pr$;
\item Calcular a matriz Google $G$;
\item Calcular a matriz de LinkRank $Lm$;
\item Calcular a matriz de Modularidade $Mm$;
\item Executar o Arrefecimento Simulado.
\end{enumerate}
\par Todos os códigos foram implementados no programa R \citep{R2010}. 
Os códigos para calcular a matriz de hiperlinks normalizada por linhas ($H$), o vetor de PageRank ($pr$) e a matriz Google ($G$) foram adaptados dos códigos disponíveis por \citet{AmyN.Langville2006}. 
O código para produzir a rede de referência \citep{Kim2010} utilizam o pacote ``Matrix" \citep{Bates2011}.

\section*{Códigos em linguagem R}
\par Código para produzir a matriz de hiperlinks normalizada por linhas ($H$):
\begin{verbatim}
rnhm <- function(A)
{
  Srow  <- rowSums(A)
  n  <- dim(A)[1]
  H  <- matrix(0,n,n)
  for (i in 1:n)
  {
    if (Srow[i] != 0)
      H[i,] <- A[i,]/Srow[i]
  }
  return(H)
}
\end{verbatim}
\par Código para calcular o vetor de PageRank ($pr$):
\begin{verbatim}
pagerank <- function(H,alpha=0.85,epsilon=1e-8,pr0)
{
  n <- dim(H)[1]
  if (missing(pr0))
    pr0 <- rep(1/n,n)
  rowsumvector <- rowSums(H) == 0
  a <- rowsumvector*1
  residual <- 1
  pr <- pr0
  while (residual >= epsilon)
  {
    prevpr <- pr
    pr <- as.vector((alpha*pr)%*%H) + (alpha*(pr%*%a)+1-alpha)*rep(1/n,n)
    residual <- dist(rbind(pr,prevpr),method='manhattan')
  }
  return(pr)
}
\end{verbatim}
\par Código para produzir a matriz Google ($G$):
\begin{verbatim}
gmatrix <- function(A,alpha=0.85)
{
  Srow <- rowSums(A)
  n  <- dim(A)[1]
  G <- matrix(0,n,n)
  for (i in 1:n)
  {
    if (Srow[i] != 0)
      G[i,] <- alpha*(A[i,]/Srow[i]) + (1/n)*(1-alpha)
    else
      G[i,] <- 1/n
  }
  return(G)
}
\end{verbatim}
\par Código para calcular a matriz de LinkRank ($Lm$):
\begin{verbatim}
linkrank <- function (G,pr)
{
  n <- length(pr)
  L <- matrix(0,n,n)
  for (i in 1:n)
    L[i,] <- pr[i] * G[i,]
  return(L)
}
\end{verbatim}
\par Código para produzir a matriz de Modularidade ($Mm$):
\begin{verbatim}
lrmM <- function(L,pr)
{
  n <- length(pr)
  qlrM <- matrix(0,n,n)
  prMA <- pr%o%pr
  for (i in 1:n)
  {
    LL <- L[i,]
    prL <- prMA[i,]
    for (j in 1:n)
    {
      qlrM[i,j] <- LL[j] - prL[j]
    }
  }
  diag(qlrM) <- 0;
  return(qlrM)
}
\end{verbatim}
\par Código do Arrefecimento Simulado:
\begin{verbatim}
kim <- function(qlrM, Tc=1, minTc=1e-10, cool=0.995, 
				max_itry, max_heat, max_rej)
{
#Parameters
  n <- dim(qlrM)[1]
  c <- 1:n
  lc <- unique(c)
  if (missing(max_itry))
    max_itry <- n^2
  if (missing(max_heat))
    max_heat <- 0.125 * max_itry
  if (missing(max_rej))
    max_rej <- max_itry
#Initial variables
  itry <- heat <- rej <- 0
  while (Tc >= minTc & rej < max_rej)
  {
    itry <- itry+1
    nsort <- sample(n,1)
    csort <- sample(lc[-which(lc==c[nsort])],1)
    LqlrM <- qlrM[nsort,]
    CqlrM <- qlrM[,nsort]
    oldm <- sum(LqlrM[c==c[nsort]]) + sum(CqlrM[c==c[nsort]])
    newm <- sum(LqlrM[c==csort]) + sum(CqlrM[c==csort])
    if (newm>=oldm)
      {c[nsort] <- csort}
    else
      {
        if (runif(1) < exp((newm-oldm)/Tc))
          {
          c[nsort] <- csort
          heat <- heat + 1
          }
        else
          {
          rej <- rej + 1
          }
      }
    if (itry > max_itry | heat > max_heat)
      {
      Tc <- cool*Tc
      itry <- heat <- rej <- 0
      }
    }
  #Last ouptut
  writeLines(paste(paste('Final Temperature =', signif(Tc,digits=1)),'\n'))
  return(c)
}
\end{verbatim}
\par Função para criar a rede de referência proposta por \citet{Kim2010}. Depende do pacote Matrix \citep{Bates2011}:
\begin{verbatim}
benchmark <- function(size,n,w=1)
{
  require(Matrix)
  comm <- Matrix(0,nrow=size, ncol=size, sparse = T)
  #internal links
  for (i in 1:(size-1))
    comm[i,i+1] <- 1
  comm[i+1,1] <- 1
  network <- comm
  for (i in 1:(n-1))
    network <- bdiag(network,comm)
  #external links
  central.node <- ceiling(median(1:size))
  for (i in 1:(n-1))
    network[central.node+size*(i-1),1+size*i] <- w
  network[central.node+(size*(n-1)),1] <- w
  return(network)
}
\end{verbatim}
\par Exemplo utilizando a rede de referência proposta por \citet{Kim2010}.
\begin{verbatim}
size <- 4; n <- 4; w <- 1
A <- benchmark(size,n,w)
H <- rnhm(A)
pr <- pagerank(H)
G <- gmatrix(A)
L <- linkrank(G,pr)
qlrM <- lrmM(L,pr)
membership <- kim(qlrM,Tc=0.1,minTc=0.0005)
\end{verbatim}

\end{document}